\documentclass[showpacs,prb,aps,twocolumn,superscriptaddress]{revtex4-1}

\pdfoutput=1
\usepackage{amsmath,amsfonts,amssymb,bm,graphicx,hyperref,color,units}
\newcommand{\W}{\mathcal{W}}
\newcommand{\llangle}{\langle\!\langle}
\newcommand{\rrangle}{\rangle\!\rangle}
\newcommand{\revision}[1]{{ #1}}

\begin{document}
\title{Distributions of electron waiting times in quantum-coherent conductors}

\author{G\'eraldine Haack}
\affiliation{D\'epartement de Physique Th\'eorique, Universit\'e de Gen\`eve, 1211 Gen\`eve, Switzerland}
\affiliation{Dahlem Center for Complex Quantum Systems and Fachbereich Physik, Freie Universit\"at Berlin, 14195 Berlin, Germany}
\author{Mathias Albert}
\affiliation{Institut Non Lin\'{e}aire de Nice, Universit\'{e} de Nice Sophia Antipolis, UMR CNRS 7335, 06560 Valbonne, France}
\author{Christian Flindt}
\affiliation{D\'epartement de Physique Th\'eorique, Universit\'e de Gen\`eve, 1211 Gen\`eve, Switzerland}
\date{\today}

\begin{abstract}
The distribution of electron waiting times is useful to characterize quantum transport in mesoscopic structures. Here we consider a generic quantum-coherent conductor consisting of a mesoscopic scatterer in a two-terminal setup. We extend earlier results for single-channel conductors to setups with several (possibly spin-degenerate) conduction channels and we discuss the effect of a finite electronic temperature.  We present detailed investigations of the electron waiting times in a quantum point contact as well as in two mesoscopic interferometers with energy-dependent transmissions: a Fabry-P\'erot interferometer and a Mach-Zehnder interferometer. We show that the waiting time distributions allow us to determine characteristic features of the scatterers, for instance the number of resonant levels in the Fabry-P\'erot interferometer that contribute to the electronic transport.
\end{abstract}

\pacs{72.70.+m, 73.23.-b, 73.63.-b}


\maketitle

\section{Introduction}

Quantum transport experiments have traditionally concerned measurements of the mean current and the shot noise in nanoscale electronic conductors. However, additional information about a quantum conductor can be obtained from the full distribution of charge transfer and its moments and cumulants.\cite{Blanter2000,Levitov1993,Levitov1996,Nazarov2003} This has led to a series of experiments measuring high-order current correlation functions in various mesoscopic structures.\cite{Reulet2003,Bylander2005,Bomze2005,Gustavsson2006,Fujisawa2006,Fricke2007,Sukhorukov2007,Timofeev2007,Gershon2008,Flindt2009,Gabelli2009,Masne2009,Fricke2010,Fricke2010b,Ubbelohde2012,Maisi2014} Typically, the distribution of transferred charge is considered on time scales that are much longer than the time intervals between subsequent charge transfers and information about short-time physics may thus be lost.

To characterize the short-time physics in a mesoscopic structure, the distribution of waiting times between electron transfers has recently been suggested as a useful quantity.\cite{Brandes2008,Welack2008,Welack2009,Albert2011,Albert2012,Thomas2013,Rajabi2013,Dasenbrook2014,Thomas2014,Wang2014,Albert2014,Sothmann2014} Waiting time distributions (WTD) are well-known in the field of classical stochastic processes\cite{Cox1962} and in quantum optics,\cite{Vyas1988,Carmichael1989} but have only recently been considered in the context of electronic transport. An elegant method to evaluate the WTD for electronic transport described by Markovian master equations has been developed by Brandes\cite{Brandes2008} and later on extended to the non-Markovian regime.\cite{Thomas2013} \revision{Within this framework, interactions can be taken into account, however, often the coupling to the leads must be treated perturbatively.} \revision{For non-interacting electrons in} coherent conductors, we have formulated a quantum theory for WTDs using scattering theory.\cite{Albert2012} This approach was recently adapted to tight-binding models\cite{Thomas2014} and generalized to periodically driven conductors using Floquet theory.\cite{Moskalets2011,Dasenbrook2014,Albert2014}

With the advent of high-frequency single-electron emitters operating in the giga-hertz regime,\cite{Feve2007,Bocquillon2013,dubois2013nature} WTDs seem particularly useful to characterize the regularity of the emitters, thereby enabling the synchronized arrival of individual electrons in a quantum electronic circuit. In terms of specific applications, WTDs have been considered for several nano-scale conductors. Coherent oscillations have been identified in the WTD for transport through double quantum dots coupled to external electrodes\cite{Brandes2008,Welack2008, Welack2009,Thomas2013,Thomas2014} and for single quantum dots coupled to one normal and one super-conducting lead.\cite{Rajabi2013} For dynamical systems, WTDs have been investigated theoretically for a periodically driven quantum capacitor \cite{Albert2011} and for clean single-particle excitations\cite{Dasenbrook2014,Albert2014} (levitons) generated by applying a sequence of Lorentzian-shaped voltage pulses to an electrode.\cite{dubois2013nature} \revision{It has been found that WTDs, unlike the shot noise and the full counting statistics of transferred charge, can clearly distinguish between electrons being emitted due to Lorentzian-shaped voltage pulses and a constant bias.\cite{Dasenbrook2014,Albert2014} For interacting quantum dot systems, effects of the Coulomb interactions between the electrons can also be identified in the WTD.\cite{Brandes2008,Thomas2013}}

\begin{figure}
  \includegraphics[width=0.9\columnwidth]{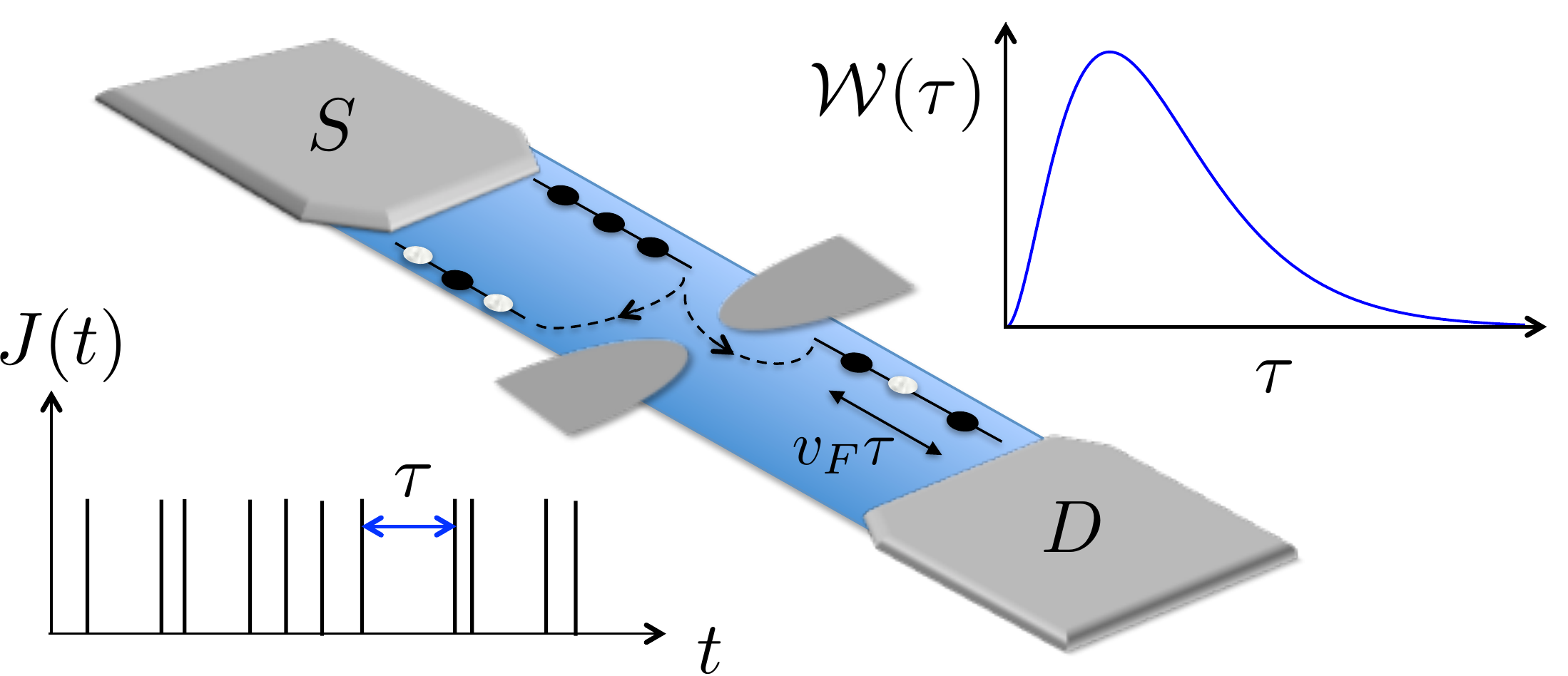}
  \caption{(Color online) A mesoscopic scatterer (here a quantum point contact) connected to source and drain electrodes. An applied voltage drives electrons through the scatterer. The distribution of waiting times $\tau$ between detection events is denoted as $\mathcal{W}(\tau)$ with $J(t)$ being the random sequence of detection events. The Fermi velocity is denoted as $v_F$.}
\label{fig:setup}
\end{figure}

The purpose of the present work is to provide a detailed account of our quantum theory of WTDs for dc-biased mesoscopic conductors\cite{Albert2012} as well as to extend the approach to setups with several conduction channels and finite electronic temperatures. To illustrate the use of our method, we consider the electronic WTD of a quantum point contact (QPC) [see Fig.~\ref{fig:setup}], a Fabry-P\'erot interferometer, and a Mach-Zehnder interferometer. Some results for the QPC were presented in Ref.~\onlinecite{Albert2012}, but here we expand considerably on our discussion of this system. The interferometers provide us with examples of scatterers which, unlike the QPC, have energy-dependent transmission amplitudes. As we will see, the WTDs allow us to determine characteristic time-scales of these scatterers.

The rest of the paper is now organized as follows. In Sec.~\ref{S2}, we provide a comprehensive account of our theory of WTDs for dc-biased conductors, including several technical steps and details that were not described in Ref.~\onlinecite{Albert2012}. We also extend our method to systems with several conduction channels and non-zero electronic temperatures. Readers who are mainly interested in the applications of our method may directly skip to Sec.~\ref{S3}, where we illustrate it with three mesoscopic scatterers.  As the first application we evaluate the electron waiting times of a biased QPC for which we, among other things, discuss the influence of the spin degree of freedom and finite electronic temperatures. We then turn to the two interferometers with energy-dependent transmission amplitudes. For the Fabry-P\'erot interferometer, we show how the number of resonances in the bias window can be identified from the WTD. For the Mach-Zehnder interferometer, we are particularly interested in the signatures of single-particle interferences which can be controlled by adjusting the path length difference of the interferometer and the applied magnetic field. Finally, in Sec.~\ref{sec:conclusions} we conclude on our work and provide a perspective on future directions and open questions. Appendix~\ref{AppenB} describes a formal analogy between one-dimensional fermions and random matrices, while in App.~\ref{AppenA} we give a brief account of renewal theory in relation to WTDs.

\section{Formalism}
\label{S2}
\subsection{Waiting time distributions}

We denote the distribution of waiting times $\tau$ between subsequent electrons as $\mathcal{W}(\tau)$. It is convenient to express the WTD in terms of the probability $\Pi(t_0,\tau+t_0)$ that no electrons are observed during the time interval \mbox{$[t_0,t_0+\tau]$} with $\tau \geq 0$. We refer to $\Pi(t_0,\tau+t_0)$ as the idle time probability. For a stationary process, the idle time probability does not depend on the reference time $t_0$, but only on the length of the interval $\tau$, such that
\begin{equation}
\Pi(t_0,\tau+t_0)=\Pi(\tau).
\end{equation}

To express the WTD in terms of the idle time probability, we choose a random time $t_0$ and consider the last observation at the earlier random time $t_e\leq t_0$. We then write the idle time probability as
\begin{equation}
\Pi(\tau)=\frac{1}{\langle \tau\rangle}\int_{-\infty}^{t_0}dt_e\left[1-\int_{t_e}^{t_0+\tau}dt \mathcal{W}(t-t_e)\right],
\label{eq:WTD2Pi}
\end{equation}
where
\begin{equation}
\langle \tau \rangle=\int_0^{\infty}\tau\W(\tau) d\tau
\end{equation}
is the mean waiting time. In Eq.~(\ref{eq:WTD2Pi}), the square brackets contain the probability that no electrons are observed in the time interval $[t_e,t_0+\tau]$. We moreover integrate over all possible times for the last electron to be observed, $-\infty\leq t_e\leq t_0$, using that the electron transfers for a stationary process are uniformly distributed in time with weight $1/\langle\tau\rangle$. Next, we change integration variables, $v=t-t_e$ and $u= \tau+t_0-t_e$, and rewrite Eq.~(\ref{eq:WTD2Pi}) as
\begin{equation}
\Pi(\tau)=\frac{1}{\langle \tau\rangle}\int^{\infty}_{\tau}du\int_{u}^{\infty}dv \mathcal{W}(v),
\label{eq:ITPcov}
\end{equation}
having used the normalization $\int_0^{\infty}d\tau\W(\tau)=1$. From this expression we see that the idle time probability indeed is independent of $t_0$. Moreover, by differentiating the idle time probability twice with respect to $\tau$ we arrive at the main result of this section
\begin{equation}
  \mathcal{W}(\tau) = \langle \tau \rangle  \frac{d^2}{d\tau^2}\Pi(\tau).
\label{eq:pi2WDT}
\end{equation}
This expression forms the basis of our further theoretical developments. We note that similar relations have found use in spectral statistics\cite{Haake2001} and quantum optics.\cite{Vyas1988}

From Eq.~(\ref{eq:pi2WDT}) we find the mean waiting time
\begin{equation}
\langle\tau\rangle = -\frac{1}{\dot{\Pi}(\tau=0)}
\label{eq:meanwaiting}
\end{equation}
by integrating over $\tau$. Here $\dot{\Pi}(\tau=0)=\frac{d}{d\tau}\Pi(\tau)|_{\tau=0}$ and we have used that $\dot{\Pi}(\tau)$ goes to zero at long times.

\subsection{Scattering states}

We consider a generic mesoscopic transport setup, where non-interacting electrons emitted from a source electrode are either transmitted through a mesoscopic conductor or are reflected back. We begin with a single conduction channel, but will later on generalize our result to a conductor with many channels. Initially, we consider a system kept at zero temperature before discussing the influence of a finite temperature of the source electrons.

To find the WTD, we use the first-quantized many-body formalism developed by Hassler \emph{et al.}\ to evaluate the idle time probability.~\cite{Hassler2008, Lesovik2011} The electronic system is brought out of equilibrium by the voltage bias $V$. We consider the electrons coming from the source electrode in the transport window
\begin{equation}
I_V=[E_F,E_F+eV].
\label{eq:trans_wind}
\end{equation}
Here we are assuming that the transmitted beam of electrons can be energetically filtered so that only electrons above the Fermi level are detected.

To describe charge transport we introduce the (left-incoming) Lippmann-Schwinger scattering states, which far from the scatterer take the asymptotic form
\begin{equation}
\varphi_k(x)=\left\{
                \begin{array}{ll}
                  e^{ikx}+r_k e^{-ikx}, & x\rightarrow - \infty \\
                  t_k e^{ikx}, & x\rightarrow \infty
                \end{array}
              \right..
\end{equation}
The reflection and transmission amplitudes are denoted as $r_k$ and $t_k$, respectively. We may also linearize the dispersion relation close to the Fermi energy as
\begin{equation}
E(k)=\hbar v_F k,
\label{eq:lin_disp}
\end{equation}
with the energy $E(k)$ and the momentum $k$ measured from the Fermi level $E_F$ and the Fermi momentum $k_F$, respectively. The Fermi velocity is
\begin{equation}
v_F=\frac{\hbar k_F}{m}
\label{eq:fermi_velo}
\end{equation}
and $m$ is the electron mass. Due to the linear dispersion, all components of a wave-packet propagate with the Fermi velocity. The scattering states are normalized as  (see e.~g.~App.~B of Ref.~\onlinecite{Choi2013})
\begin{equation}
\int_{-\infty}^{\infty} dx \,\, \varphi^{*}_{k'}(x)\varphi_{k}(x)=2\pi\delta(k-k').
\label{eq:lippmann_ortho}
\end{equation}

The scattering states, however, are inconvenient to work with due to this unusual normalization. To regularize the problem, we divide the transport window $I_V$ into $N$ energy compartments of size $eV/N$ with corresponding momentum intervals of size
\begin{equation}
\kappa=\frac{eV}{N\hbar v_F}.
\label{eq:kappa}
\end{equation}
For each compartment, $m=1,\ldots, N$, we define the orthonormalized single-particle states
\begin{equation}
\psi_m(x,t) =\langle x|\psi_m(t)\rangle=\frac{1}{\sqrt{2\pi\kappa}}\int_{(m-1)\kappa}^{m\kappa} \!\!\!\!\!\!\!\! dk\, e^{-iv_F k t} \varphi_k(x),
\label{eq:sing_part_stat}
\end{equation}
where we have explicitly included the time-dependence. These states are appropriately orthonormalized
\begin{equation}
\langle\psi_m(t)|\psi_n(t)\rangle=\int_{-\infty}^{\infty} dx \,\, \psi^{*}_{m}(x,t)\psi_{n}(x,t)=\delta_{nm}
\end{equation}
as it follows using Eq.~(\ref{eq:lippmann_ortho}). From the single-particle states in Eq.~(\ref{eq:sing_part_stat}) we build the $N$-particle many-body state expressed by the Slater determinant
\begin{equation}
|\Psi_S^{(N)}\rangle=\frac{1}{\sqrt{N!}}\sum_{\pi\in S_N}\mathrm{sgn}(\pi)\,|\psi_{\pi(1)}\rangle \otimes\cdots\otimes |\psi_{\pi(N)}\rangle.
\label{eq:Slater}
\end{equation}
Here $\pi$ is an element of the symmetric group $S_N$ of $N$ elements and $\mathrm{sgn}(\pi)=\pm1$ is the sign of the permutation. Next, we use the  $N$-particle Slater determinant to obtain the idle time probability. To mimic a stationary process, we eventually take the limit of many particles, $N\gg1 $.

\subsection{Determinant formula}

To derive the idle time probability from scattering theory we make use of the linear dispersion relation. Instead of considering the probability of detecting no charges in the \emph{temporal} interval $[t_0,t_0+\tau]$ at the position $x_0$ after the scatterer, we fix the time and consider the probability of detecting no charges in the  \emph{spatial} interval $[x_0-v_F\tau,x_0]$. We can define an operator, which projects single-particle states onto this line segment
\begin{equation}
\mathcal{\widehat{Q}}_{\tau}=\int_{x_0-v_F\tau}^{x_0}\! \! \! \! \! \!  dx\,|x\rangle\!\langle x|.
\end{equation}
The expectation value of $1-\mathcal{\widehat{Q}}_{\tau}$ with respect to a single-particle state is then the probability of \emph{not} finding a given particle in the spatial region. For $N$ particles, the operator $1-\mathcal{\widehat{Q}}_{\tau}$ should be applied to each occupied single-particle state, such that the idle time probability corresponding to the Slater determinant in Eq.~(\ref{eq:Slater}) becomes
\begin{equation}
\Pi(\tau)=\langle \Psi_S^{(N)}(\tau)|\bigotimes_{i=1}^{N}(1-\mathcal{\widehat{Q}}_{\tau})|\Psi_S^{(N)}(\tau)\rangle.
\end{equation}
Using the expression for the Slater determinant, we find
\begin{equation}
\Pi(\tau)= \frac{1}{N!}\sum_{\pi,\pi'\in S_N}\!\!\!\!\!\mathrm{sgn}(\pi\circ\pi')\prod_{i=1}^N \langle\psi_{\pi(i)}|  (1-\mathcal{\widehat{Q}})|\psi_{\pi'(i)}\rangle.
\end{equation}
since $\mathrm{sgn}(\pi)\mathrm{sgn}(\pi')=\mathrm{sgn}(\pi\circ\pi')$, where $\pi\circ\pi'=\pi''$ is an element in $S_N$. Rearranging the expression, we obtain
\begin{equation}
\begin{split}
\Pi(\tau)&= \frac{1}{N!}\sum_{\pi,\pi''\in S_N}\!\!\!\!\!\mathrm{sgn}(\pi'')\prod_{i=1}^N \langle\psi_{i}|  (1-\mathcal{\widehat{Q}})|\psi_{\pi''(i)}\rangle\\
&= \sum_{\pi''\in S_N}\!\!\!\!\!\mathrm{sgn}(\pi'')\prod_{i=1}^N \langle\psi_{i}|  (1-\mathcal{\widehat{Q}}_{\tau})|\psi_{\pi''(i)}\rangle\\
\end{split}
\end{equation}
and finally
\begin{equation}
\Pi(\tau)=\det(1-\mathbf{Q}_\tau).
\label{eq:determintant}
\end{equation}
Here, the single-particle matrix elements of $\mathbf{Q}_\tau$ are
\begin{equation}
\begin{split}
[\bold{Q}_\tau]_{m,n} &= \langle\psi_m(\tau)|\mathcal{\widehat{Q}}_\tau|\psi_n(\tau)\rangle\\
&= \int_{(m-1)\kappa}^{m\kappa}  \int_{(n-1)\kappa}^{n\kappa} \!\! \frac{dk' dk}{2\pi\kappa} \,\, t^*_{k'} t_k  \, K_\tau (k-k'),
\end{split}
\label{eq:QelementsPre}
\end{equation}
and the kernel is
\begin{equation}
K_\tau(q) =  \frac{2e^{- i q v_F \tau/2}\sin(\frac{q v_F \tau}{2})}{q} .
\end{equation}
We note that the kernel is closely related to the coherence function of single electrons emitted by a dc-source.\cite{Haack2011,Haack2013} Taking the limit $N\gg 1$, we get $\kappa\ll 1$ from Eq.~(\ref{eq:kappa}), and we can approximate the matrix elements as
\begin{equation}
[\bold{Q}_{\tau}]_{m,n} \simeq \frac{\kappa t^*_{\kappa m}t_{\kappa n}}{2\pi}  K_\tau(\kappa n-\kappa m).
\label{eq:Qelements}
\end{equation}

Combining Eqs.\ \eqref{eq:pi2WDT}, \eqref{eq:determintant}, and \eqref{eq:Qelements}, we may now calculate the WTD for an arbitrary dc-scattering problem. All details about the scatterer enter via the transmission amplitudes in Eq.~\eqref{eq:Qelements}. Equations \eqref{eq:pi2WDT} and \eqref{eq:determintant} can moreover be combined using Jacobi's formula, which expresses the derivative of a determinant of a matrix $\mathbf{A}$ as
\begin{equation}
\frac{d}{d\tau}\det(\mathbf{A})=\mathrm{Tr}\{\mathrm{adj}(\mathbf{A})\frac{d}{d\tau}\mathbf{A}\}.
\end{equation}
Here, the adjugate of $\mathbf{A}$ is denoted as $\mathrm{adj}(\mathbf{A})$. For an invertible matrix, $\mathrm{adj}(\mathbf{A})=\det(\mathbf{A})\mathbf{A}^{-1}$.

As a first application of Jacobi's formula, we evaluate the mean waiting time from Eq.\ \eqref{eq:meanwaiting}, taking $\mathbf{A}=1-\mathbf{Q}_\tau$,
\begin{equation}
\langle\tau\rangle = \frac{1}{\mathrm{Tr}\{\mathbf{\dot{Q}}_0\}}.
\label{eq:meanwaiting2}
\end{equation}
Above, we have defined $\mathbf{\dot{Q}}_0=\frac{d}{d\tau}\mathbf{Q}_\tau|_{\tau=0}$ and used that $\Pi(\tau=0)=1$. In general, $\Pi(\tau)=\det(1-\mathbf{Q}_\tau)>0$ at any finite $\tau$, such that $1-\mathbf{Q}_\tau$ is invertible and
\begin{equation}
\mathrm{adj}(1-\mathbf{Q}_\tau)=\Pi(\tau)\mathbf{g}_\tau,
\end{equation}
having defined
\begin{equation}
\mathbf{g}_\tau=(1-\mathbf{Q}_\tau)^{-1}.
\label{eq:resolvent}
\end{equation}
From Eq.~\eqref{eq:Qelements} we easily find
\begin{equation}
\mathrm{Tr}\{\mathbf{\dot{Q}}_0\}=\mathrm{Tr}\{\mathbf{\dot{Q}}_\tau\} =\frac{eV}{h}\sum_{n=1}^N\frac{|t_{\kappa n}|^2}{N},
\end{equation}
which is just the mean particle current. For a fully transmitting scatterer with $|t_k|^2=1$ we obtain
\begin{equation}
\langle\tau\rangle = \frac{h}{eV} \equiv \bar{\tau}.
\label{eq:meanwaiting3}
\end{equation}
This result shows that the mean waiting time between the incoming electrons is simply $\bar{\tau}=h/eV$. The applied voltage bias $V$ enters only through the mean waiting time between the incoming electrons. This is correct as long as the voltage is much smaller than the Fermi energy, $eV\ll E_F$, such that the dispersion relation remains linear. Additionally, the mean waiting time together with the Fermi velocity defines a length scale
\begin{equation}
\ell=\bar{\tau}v_F=\lambda_F\frac{E_F}{eV},
\end{equation}
where $\lambda_F =2\pi/k_F$ is the Fermi wavelength. To investigate quantum properties of the incoming state and the coherence properties of the scatterer, the length scale $\ell$ should be smaller than the coherence length of the sample. This condition is typically fulfilled in mesoscopic samples.

For the WTD, we find after some algebra
\begin{equation}
  \mathcal{W}(\tau) = \frac{\Pi(\tau)}{\mathrm{Tr}\{\dot{\mathbf{Q}}_0\}}\left[\mathrm{Tr}^2\{\mathbf{g}_\tau \dot{\mathbf{Q}}_\tau\}-\mathrm{Tr}\{(\mathbf{g}_\tau \dot{\mathbf{Q}}_\tau)^2+\mathbf{g}_\tau \ddot{\mathbf{Q}}_\tau\}\right].
\label{eq:finalWTD}
\end{equation}
This expression can be evaluated for an arbitrary time-independent scattering problem and constitutes the central result of this section.

\subsection{Finite temperatures}

We now discuss the case where the incoming electrons are emitted from a source electrode at a finite temperature, while the drain electrode is kept at zero temperature. The Fermi-Dirac distribution of the electrons occupying the incoming states is
\begin{equation}
f_\beta(\varepsilon)=\frac{1}{e^{\beta(\varepsilon-E_F-eV)}+1},
\end{equation}
where $\beta =1/k_BT$ is the inverse temperature. (For the rest of the paper, we use $\beta$ when referring to temperature, while $T$ denotes the transmission probability.) To describe the effect of a finite electronic temperature, we extend the energy window in Eq.~(\ref{eq:trans_wind}) to
\begin{equation}
I_\beta=[E_F,E_F+E_\beta],
\end{equation}
where the cut-off $E_\beta>eV$ is chosen such that $f_\beta(E_\beta+E_F)\simeq 0$. Similarly to the zero-temperature case, we divide the energy window into $N$ compartments of size $E_\beta/N$ with corresponding momentum intervals of size
\begin{equation}
\kappa_\beta=\frac{E_\beta}{N\hbar v_F}.
\end{equation}
For the matrix elements of $\mathbf{Q}_\tau$ we take (see also Ref.~\onlinecite{Hassler2008})
\begin{equation}
[\bold{Q}_{\tau}]_{m,n}\! =\!  \int_{(n-1)\kappa_\beta}^{n\kappa_\beta}\int_{(m-1)\kappa_\beta}^{m\kappa_\beta}  \!\! \frac{dk' dk}{2\pi\kappa_\beta} g(k')g(k)t^*_{k'} t_k  \, K_\tau (k-k')
\label{eq:Qelements_finiteT}
\end{equation}
with
\begin{equation}
\label{eq:g}
g(k)=\sqrt{f_\beta(\hbar v_F k)}.
\end{equation}
At zero temperature, this expression reduces to Eq.~(\ref{eq:QelementsPre}). Moreover, in the limit of many particles, $N\gg1$, we find
\begin{equation}
[\bold{Q}_{\tau}]_{m,n} \simeq  \frac{\kappa_\beta t^*_{\kappa m}t_{\kappa_\beta n}g(\kappa_\beta m)g(\kappa_\beta n)}{2\pi}  K_\tau(\kappa_\beta n-\kappa_\beta m)
\label{eq:QelementsT}
\nonumber
\end{equation}
as an extension of the zero-temperature result in Eq.~(\ref{eq:Qelements}).

\subsection{Several channels}

We now generalize our theory to $M$ identical conduction channels. Assuming that the channels are independent, we may write the idle time probability for the $M$ channels as
\begin{equation}
\Pi_M(\tau)=[\Pi_1(\tau)]^M
\label{eq:idletime_Nchannel}
\end{equation}
where $\Pi_1(\tau)$ is the idle time probability of a single channel. The mean waiting time and the WTD then become
\begin{equation}
\langle\tau\rangle_M=\frac{\langle\tau\rangle_1}{M}
\label{eq:meanwait_Nchannel}
\end{equation}
and
\begin{equation}
  \mathcal{W}_M(\tau) = \frac{[\Pi_1(\tau)]^M}{\mathrm{Tr}\{\dot{\mathbf{Q}}_0\}}\!\left[M\mathrm{Tr}^2\{\mathbf{g}_\tau \dot{\mathbf{Q}}_\tau\}\!-\!\mathrm{Tr}\{(\mathbf{g}_\tau \dot{\mathbf{Q}}_\tau)^2\!+\!\mathbf{g}_\tau \ddot{\mathbf{Q}}_\tau\}\right],
\label{eq:finalWTD_Nchannel}
\end{equation}
where $\langle\tau\rangle_1$ is the mean waiting time of a single quantum channel. This expression can be evaluated for an arbitrary scattering problem with $M$ independent channels. For example, a system with two identical, independent spin channels would correspond to $M=2$.

It is instructive to consider the limit of many channels, $M\gg 1$. In that case, the first term of Eq.\ \eqref{eq:finalWTD_Nchannel} dominates and we may write
\begin{equation}
  \mathcal{W}_M(\tau) \simeq \frac{M[\Pi_1(\tau)]^M}{\mathrm{Tr}\{\dot{\mathbf{Q}}_0\}}\mathrm{Tr}^2\{\mathbf{g}_\tau \dot{\mathbf{Q}}_\tau\}.
\end{equation}
The idle time probability $\Pi_1(\tau)$ is unity at $\tau=0$ and decays to zero at long times. With many channels, $[\Pi_1(\tau)]^M$ is only on the order of unity for $\tau\ll\langle\tau\rangle_1$ and then decays rapidly to zero at longer times. We can therefore use the short-time approximation $\mathbf{g}_\tau\simeq 1$ and write
\begin{equation}
\begin{split}
  \mathcal{W}_M(\tau) &\simeq M\mathrm{Tr}\{\dot{\mathbf{Q}}_0\}[\Pi_1(\tau)]^M\\
    &= M\mathrm{Tr}\{\dot{\mathbf{Q}}_0\}e^{M\log\{\Pi_1(\tau)\}},
\end{split}
\end{equation}
having used $\mathrm{Tr}\{\dot{\mathbf{Q}}_\tau\}=\mathrm{Tr}\{\dot{\mathbf{Q}}_0\}$ in the first line. Additionally, we rewrite $\log\{\Pi_1(\tau)\}=\log\{\det(1-\mathbf{Q}_\tau)\}=\mathrm{Tr}\{\log(1-\mathbf{Q}_\tau)\}$, which for $\tau\ll\langle\tau\rangle_1$ can be expanded as $\mathrm{Tr}\{\log(1-\mathbf{Q}_\tau)\}\simeq -\mathrm{Tr}\{\mathbf{Q}_\tau\}=-\mathrm{Tr}\{\dot{\mathbf{Q}}_0\}\tau=-\tau/\langle\tau\rangle_1$. We then finally obtain
\begin{equation}
  \mathcal{W}_M(\tau) \simeq  \frac{e^{-\tau/\langle\tau\rangle_M}}{\langle\tau\rangle_M}, \,\,\, M\gg 1.
\end{equation}
Thus, as the number of channels is increased, the WTD approaches an exponential distribution with rate $1/\langle\tau\rangle_M$. The WTD would also be exponential with a large number of non-identical, but independent channels.

\section{Applications}
\label{S3}

We are now ready to illustrate our method with several examples of mesoscopic scatterers. \revision{These examples illustrate how WTDs contain information about the quantum transport which is complementary to what can be learned from the shot noise and the full counting statistics of transferred charge.}

\subsection{Quantum point contact}

We first consider a quantum point contact (QPC) \revision{with an approximately energy-independent transmission $|t_k|^2=T$ in the transport window.\cite{Buttiker1990}} To begin with, the temperature is set to zero and the electrons are considered spinless. Later on we include finite-temperature effects and the spin degree of freedom.

\begin{figure}
\includegraphics[width=0.95\linewidth]{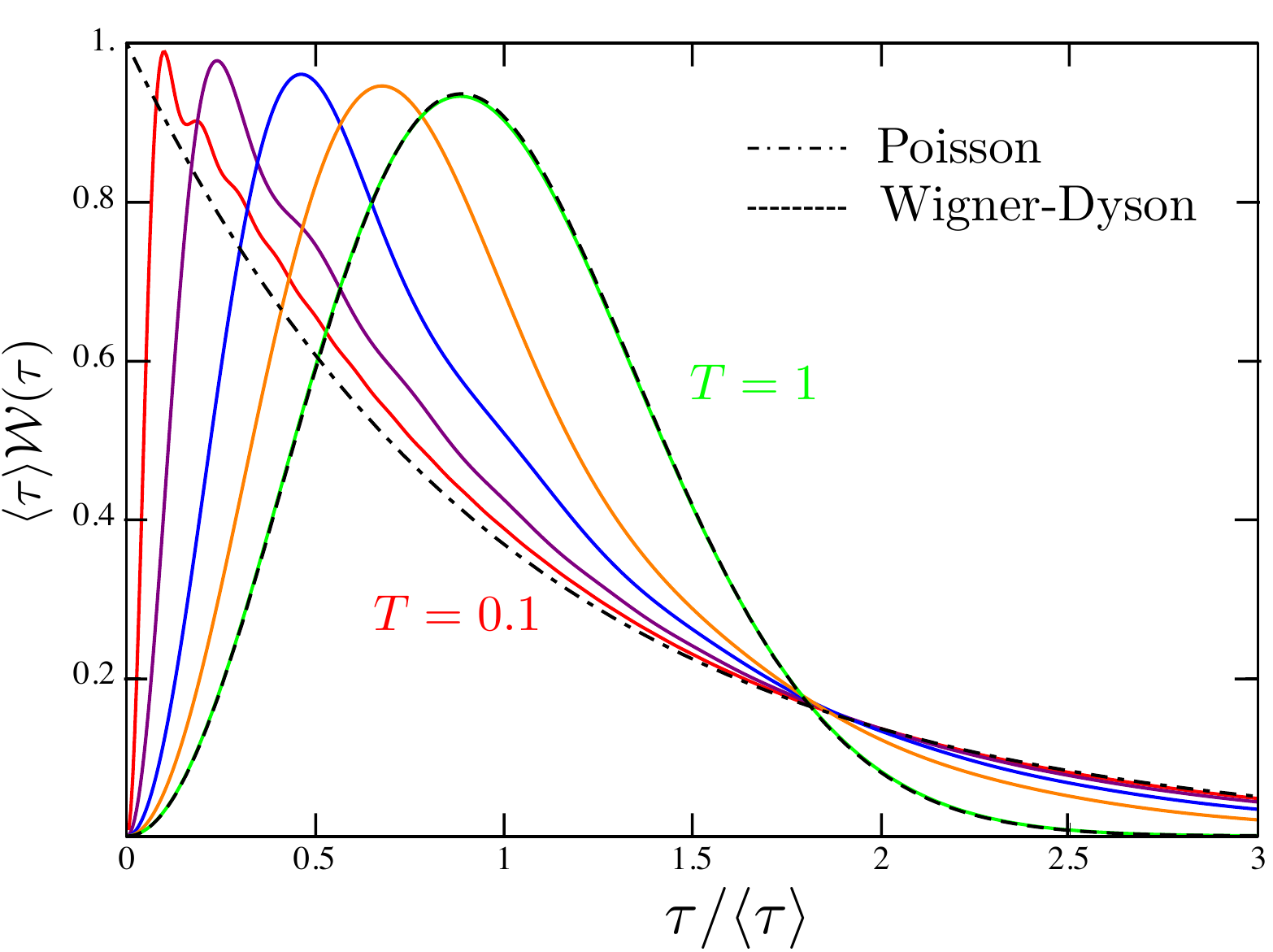}
\caption{(Color online)  WTD for a QPC with transmission $T=0.1$, 0.25, 0.5, 0.75, 1. For $T=1$, the WTD is well-approximated by the Wigner-Dyson distribution in Eq.~(\ref{eq:wigner}). Close to pinch-off ($T\simeq 0$), the WTD approaches Poisson statistics given by the exponential distribution $\mathcal{W}(\tau)\simeq e^{-\tau/\langle\tau\rangle}/\langle\tau\rangle$. For low transmissions, small oscillations with period $\bar{\tau}$ are superimposed on the exponentially decaying WTD.   \label{fig:wtd}}
\end{figure}

Figure \ref{fig:wtd} shows WTDs for different values of the transmission $T$. As we will discuss in detail, the WTDs exhibit a crossover from Wigner-Dyson statistics at full transmission ($T=1$) to Poisson statistics close to pinch-off ($T\simeq0$). At $T=1$, the quantum channel is fully open and there is no partition noise due to the QPC. All fluctuations then arise from the quantum statistics encoded in the incoming many-body electronic state. In this case, the WTD consists of a single peak centered around $\tau=\bar{\tau}$. If the electron transport was purely deterministic, the WTD would be a Dirac peak at $\tau=\bar{\tau}$. However, as we find, large fluctuations around the mean value reflect the uncertainty associated with the wave nature of the electrons. The correlations between the electrons are due to the Pauli principle only which ensures that two electrons cannot occupy the same state. This is reflected in the suppression of the WTD at $\tau=0$. This suppression is similar to the Fermi-hole in the density-density correlation function of a free electron gas. \cite{Landau1959, Buttiker1992} The fermionic correlations also force the WTD to decay strongly beyond a few mean waiting times, where it essentially vanishes.

To understand the shape of the WTD at full transmission, we use a mapping between one-dimensional fermions and random matrix theory. In Appendix \ref{AppenB}, we show how to map the many-body wave function of non-interacting fermions to the joint probability distribution of eigenvalues of random matrices. In general, the canonical ensembles of random matrices, labeled by their symmetry parameter $\beta$, can be mapped onto the Calogero-Sutherland model of one-dimensional fermions with coupling constant proportional to $\beta-2$.\cite{Calogero1969, Sutherland1971} Free fermions correspond to $\beta=2$ if we replace the $N$ spatial coordinates by the $N$ eigenvalues of a random matrix of the Gaussian unitary ensemble. As a direct consequence, all spatial correlation functions of the original problem are formally given by the spectral correlation functions of the corresponding random matrix theory problem. Additionally, since the dispersion relation is linear, the system is invariant under Galilean transformation and the spatial correlations are identical to the temporal correlations up to a scaling factor (here the Fermi velocity). We can then use a result from random matrix theory to express the WTD at full transmission as
\begin{equation}
  \mathcal W^{\mathrm{WD}}(\tau)=\frac{32\tau^2}{\pi^2\bar{\tau}^3}e^{-4\tau^2/\pi\bar{\tau}^2}  \,.
  \label{eq:wigner}
\end{equation}
This distribution is known as the Wigner-Dyson surmise in the context of spectral statistics. In Fig.~\ref{fig:wtd}, we show that the Wigner-Dyson distribution is in very good agreement with our exact results.

Next, we turn to the discussion of the WTD as the transmission $T$ is reduced below unity. The QPC now introduces partition noise as electrons may reflect back on the QPC, allowing for longer waiting times between transferred charges. The mean waiting time for non-perfect transmission is
\begin{equation}
\langle\tau\rangle=\frac{\bar{\tau}}{T}.
\end{equation}
The Fermi-hole persists independently of the transmission as it follows from a short-time expansion of the WTD
\begin{equation}
  \mathcal W(\tau)\simeq \frac{\pi^2}{3}\frac{T}{\bar{\tau}}\left(\frac{\tau}{\bar{\tau}}\right)^2\,,\quad \tau\ll\bar{\tau}\,.
  \label{eq:wtd_short_times}
\end{equation}
We see how the Pauli exclusion principle manifests itself in a complete suppression at $\tau=0$ for all values of the transmission. The quadratic dependence at short times is a generic behavior as long as the energy spectrum can be linearized and the wave-functions are differentiable.\cite{Albert2014} As the QPC comes close to pinch-off, the transmissions become rare and uncorrelated corresponding to a Poisson process. The WTD can then be approximated by an exponential distribution as indicated in Fig.~\ref{fig:wtd}.

\begin{figure}
  \includegraphics[width=0.95\columnwidth]{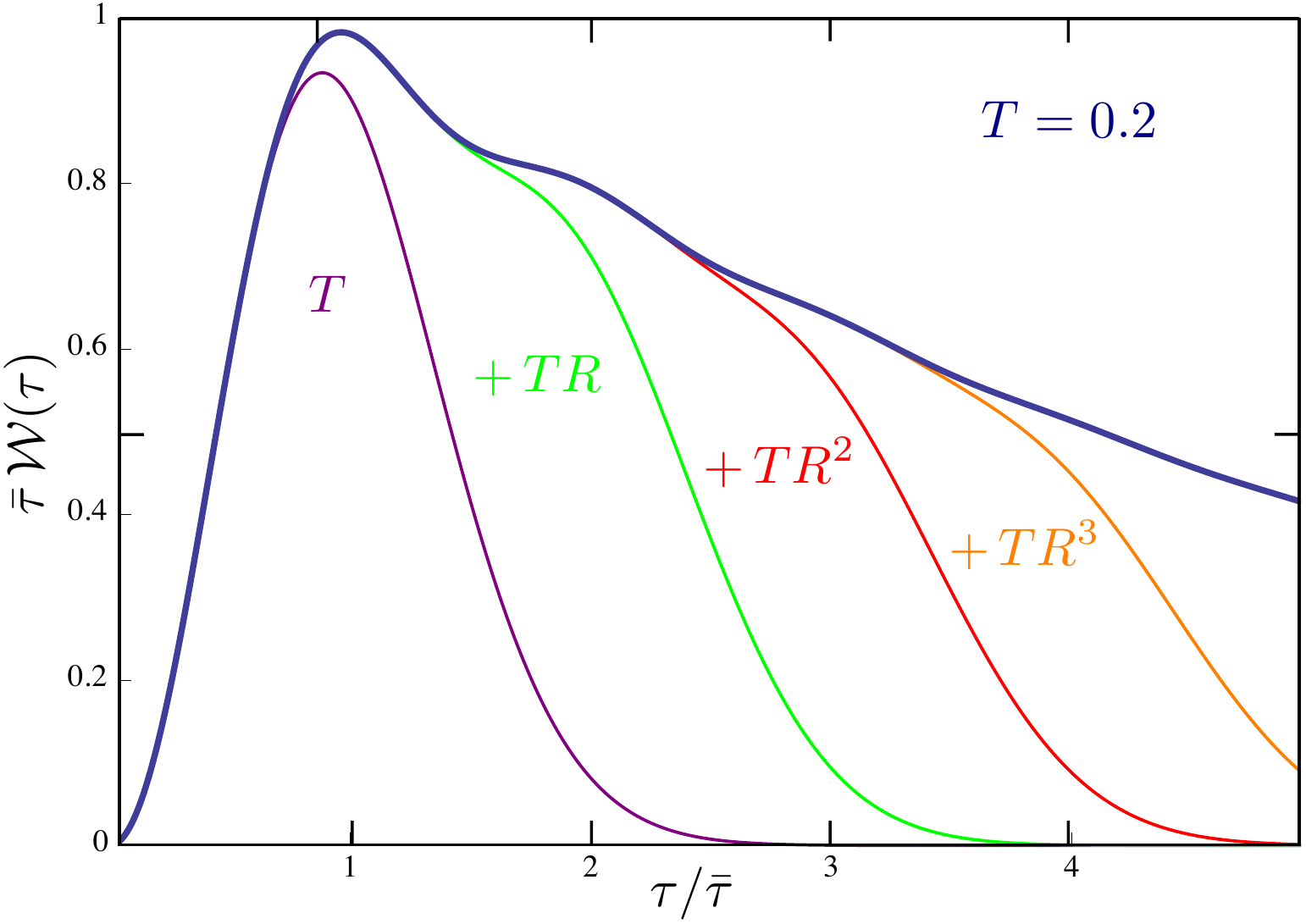}
  \caption{(Color online) Expansion of the WTD in the number of reflections that have occurred. The thick line shows the exact result, while the thin lines indicate the series in Eq.~(\ref{eq:Rexpansion}) as an increasing number of terms are included.}
  \label{fig3}
\end{figure}

In Ref.~\onlinecite{Albert2012}, it was found that the crossover from Wigner-Dyson statistics to Poisson statistics is accompanied by small oscillations on top of the exponentially decaying WTD. The period of the oscillations approaches $\bar{\tau}$ in the limit of a vanishing transmission. To understand the WTD with a transmission below unity, we resolve it with respect to the number of reflections that have occurred (see Fig.~\ref{fig3} together with the explanation below). Formally, we expand the WTD as
\begin{equation}
\mathcal{W}(\tau)=T\mathcal{W}^{(0)}(\tau)+TR\mathcal{W}^{(1)}(\tau)+TR^2\mathcal{W}^{(2)}(\tau)+\ldots,
\label{eq:Rexpansion}
\end{equation}
where $R=1-T$ is the reflection probability and $\mathcal{W}^{(n)}(\tau)$ is the WTD given that $n$ reflections have occurred during the waiting time $\tau$. In Ref.~\onlinecite{Dasenbrook2014}, such $n$-resolved WTDs were calculated for Lorentzian-shaped voltage pulses using a renewal assumption (see Appendix~\ref{AppenA} for a discussion of renewal theory). Here, in contrast, we evaluate them exactly by explicitly expanding the WTD in $R$. To this end, we rewrite Eq.~(\ref{eq:pi2WDT}) for the WTD as
\begin{equation}
\begin{split}
\mathcal{W}(\tau)&=\frac{T\bar{\tau}}{(1-R)^2}\frac{d^2}{d\tau^2}\Pi(\tau)\\
&=T\bar{\tau}\sum_{n=0}^\infty (n+1)R^n \frac{d^2}{d\tau^2}\sum_{m=0}^\infty\frac{R^m}{m!}\Pi^{(m)}(\tau),
\end{split}
\end{equation}
where $\Pi^{(m)}(\tau)=\partial_R^m\Pi(\tau)|_{R=0}$. Collecting terms to same order in $R$, we can identify the $n$-resolved WTDs in Eq.~(\ref{eq:Rexpansion}). For the first two WTDs, we find
\begin{equation}
\begin{split}
\mathcal{W}^{(0)}(\tau)&=\bar{\tau}\frac{d^2}{d\tau^2}\Pi^{(0)}(\tau),\\
\mathcal{W}^{(1)}(\tau)&=\bar{\tau}\frac{d^2}{d\tau^2}\left(2\Pi^{(0)}(\tau)+\Pi^{(1)}(\tau)\right),\\
\end{split}
\end{equation}
Moreover, by using Jacobi's formula we obtain
\begin{equation}
\begin{split}
\Pi^{(0)}(\tau)=&\det(1-\mathbf{Q}^{(0)}_\tau),\\
\Pi^{(1)}(\tau)=&\Pi^{(0)}(\tau)\mathrm{Tr}\{\mathbf{g}^{(0)}_\tau\mathbf{Q}^{(0)}_\tau\},\\
\end{split}
\end{equation}
where
\begin{equation}
[\bold{Q}_{\tau}^{(0)}]_{m,n} \simeq \frac{ \kappa }{2\pi} K_\tau(\kappa n-\kappa m)
\end{equation}
and
\begin{equation}
\mathbf{g}^{(0)}_\tau=(1-\mathbf{Q}^{(0)}_\tau)^{-1}
\end{equation}
correspond to the fully transmitting QPC. The $n$-resolved WTDs for larger $n$ are found in a similar way.

In Fig.~\ref{fig3}, we show an expansion of the WTD in the number of reflections that have occurred. At short times, it is unlikely that any reflections occur and the WTD is well-approximated by a Wigner-Dyson distribution with a peak around $\bar{\tau}$. For longer times, higher-order WTDs become important and we need to include more terms in Eq.~(\ref{eq:Rexpansion}) corresponding to several reflections. These terms give rise to peaks at higher multiples of $\bar{\tau}$. It is worth noting that the influence of a finite QPC transmission is formally equivalent to the effect of a finite detector efficiency on the spectral statistics of complex systems like nuclei as investigated by Pato and Bohigas.\cite{Bohigas2004, Bohigas2006} Starting from a complete \revision{random matrix theory} spectrum and randomly removing energy levels they showed that the level spacing distribution exhibits a crossover from a Wigner-Dyson distribution to Poisson statistics as an increasing number of levels are removed. In the case of the QPC, the random process of removing levels corresponds to the random reflection (or removal) of electrons from the electronic many-body state.

\begin{figure}
  \includegraphics[width=0.94\columnwidth]{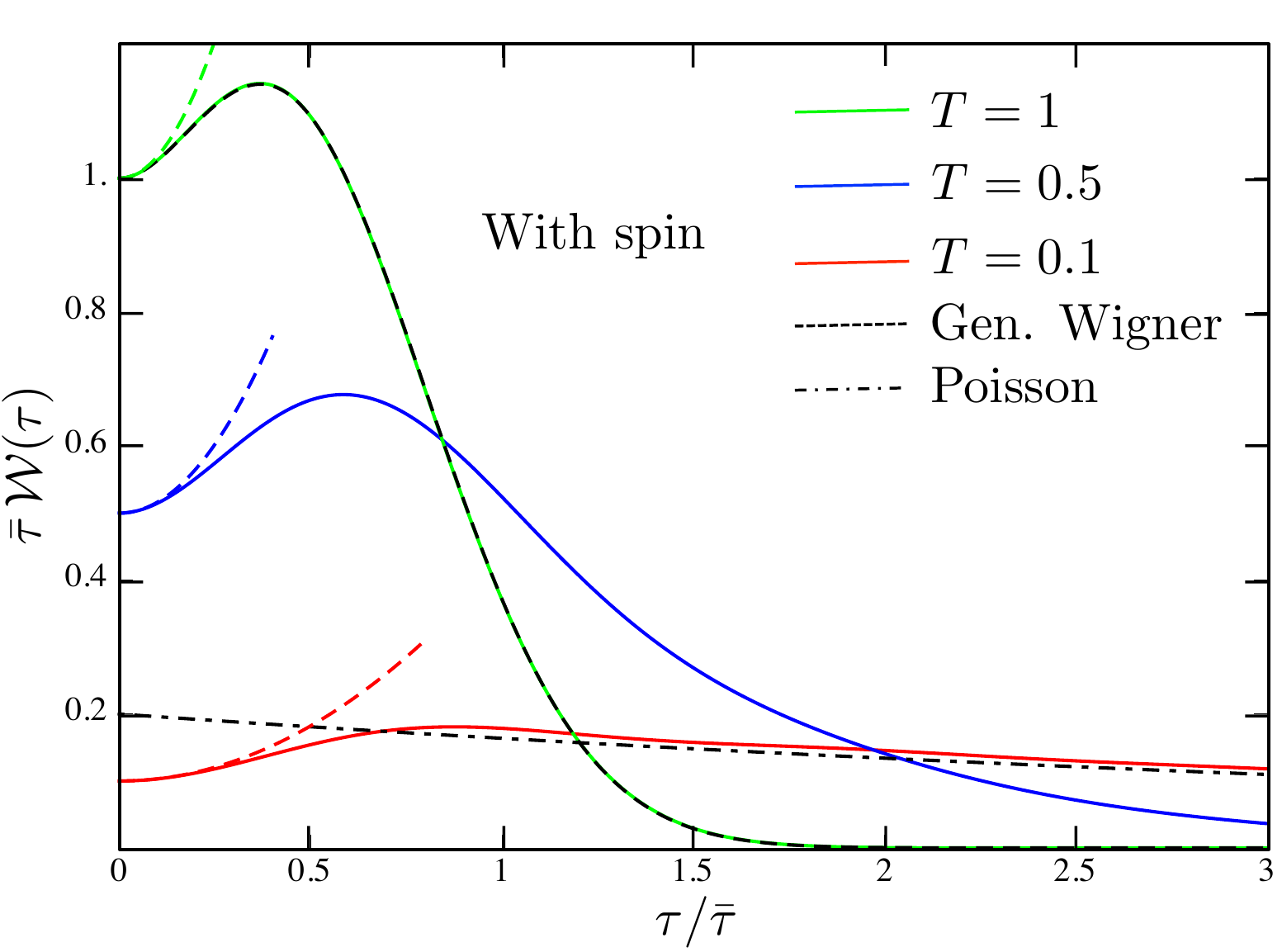}
  \caption{(Color online) WTD of a single-channel QPC including spin. At full transmission ($T=1$) the WTD is well-described by the generalized Wigner-Dyson distribution in Eq.~(\ref{eq:wigner2}). Close to pinch-off ($T\simeq 0$), the WTD approaches Poisson statistics given by the exponential distribution $\mathcal{W}(\tau)\simeq e^{-\tau/\langle\tau\rangle}/\langle\tau\rangle$. The colored dashed lines indicate the short-time expansion given by Eq.~(\ref{eq:wtd_short_times_spin}). \label{fig:wtdspin1}}
\end{figure}

We now include spin by taking $M=2$ in Eq.~(\ref{eq:finalWTD_Nchannel}). Figure~\ref{fig:wtdspin1} shows results for the WTD for different values of the transmission $T$. For the fully open QPC, we can generalize the Wigner surmise by combining Eqs.~(\ref{eq:idletime_Nchannel},\ref{eq:meanwait_Nchannel}) and (\ref{eq:wigner}). We then find
\begin{equation}
\label{eq:wigner2}
   \mathcal W_2^{\mathrm{WD}}(\tau)=\bar{\tau}[\dot{\Pi}^{\mathrm{WD}}(\tau)]^2+\Pi^{\mathrm{WD}}(\tau) \mathcal W^{\mathrm{WD}}(\tau),
\end{equation}
where
\begin{equation}
\Pi^{\mathrm{WD}}(\tau)=e^{-4\tau^2/\pi\bar{\tau}^2}-(\tau/\bar{\tau})\textrm{erfc}(2 \tau/\sqrt{\pi}\bar{\tau})
\end{equation}
follows from Eq.~(\ref{eq:ITPcov}) and $\textrm{erfc}(x)$ is the complementary error function. It can be shown that the mean waiting time corresponding to this distribution indeed is $\langle \tau \rangle=\bar{\tau}/2$ in agreement with Eq.~(\ref{eq:meanwait_Nchannel}).

Figure~\ref{fig:wtdspin1} shows that the generalized Wigner-Dyson distribution in Eq.~(\ref{eq:wigner2}) is in good agreement with our exact results for the fully transmitting QPC. In addition, as the transmission $T$ approaches zero, the WTD is again well-approximated by an exponential distribution with mean waiting time $\langle \tau \rangle=\bar{\tau}/2T$. The  short-time expansion of the WTD including spin reads
\begin{equation}
  \mathcal W_2(\tau)\simeq \frac{T}{\bar{\tau}}+\frac{\pi^2 }{3}\frac{T}{\bar{\tau}}\left(\frac{\tau}{\bar{\tau}}\right)^2\,,\quad \tau\ll\bar{\tau},
  \label{eq:wtd_short_times_spin}
\end{equation}
and is also indicated in Fig.~\ref{fig:wtdspin1}. Two electrons can now be detected at the same time, since the Pauli principle does not prevent two electrons with different spin to be in the same orbital state. The Fermi hole around $\tau=0$ is then lifted as we see. \revision{The results with and without spin clearly illustrate how the WTD contains information about the fermionic nature of electrons.}

\begin{figure}
  \includegraphics[width=0.95\columnwidth]{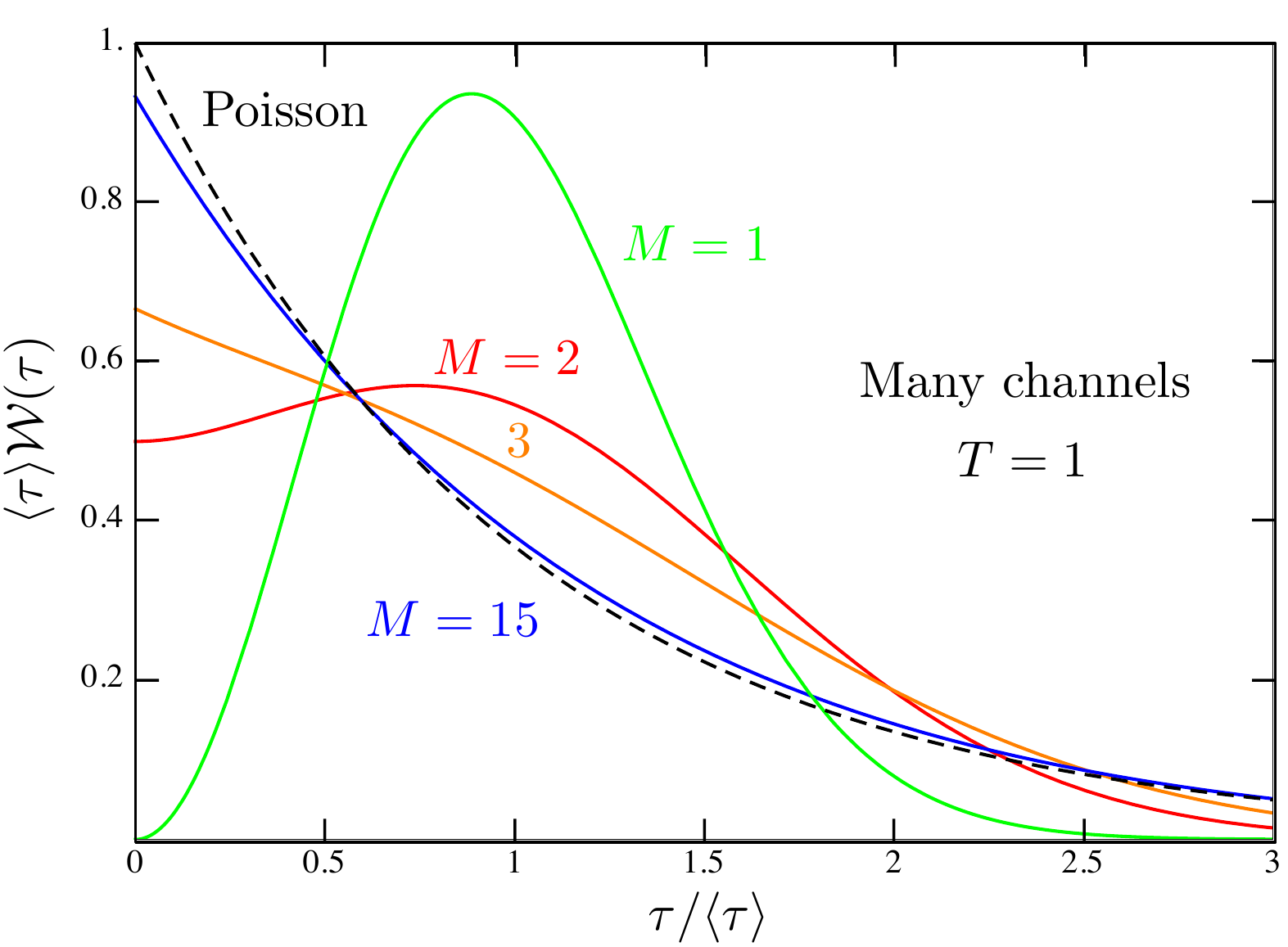}
  \caption{(Color online) WTDs with an increasing number of conduction channels. As the number of fully transmitting channels is increased from $M=1$ to $M=15$, we see a crossover from Wigner-Dyson statistics to Poisson statistics. The suppression of the WTD at short times is completely lifted as the number of channels is increased.}
  \label{fig:many_channels}
\end{figure}

In Fig.~\ref{fig:many_channels} we increase the number of open conduction channels. With $M=1$ and $M=2$ fully open channels we recover the results for a single-channel QPC with and without spin. As the number of channels is increased, the WTD exhibits a crossover to Poisson statistics with an exponential distribution of waiting times. In this case, the suppression of the WTD at short times is completely lifted, since the conduction channels are independent.

\begin{figure}
  \includegraphics[width=0.95\columnwidth]{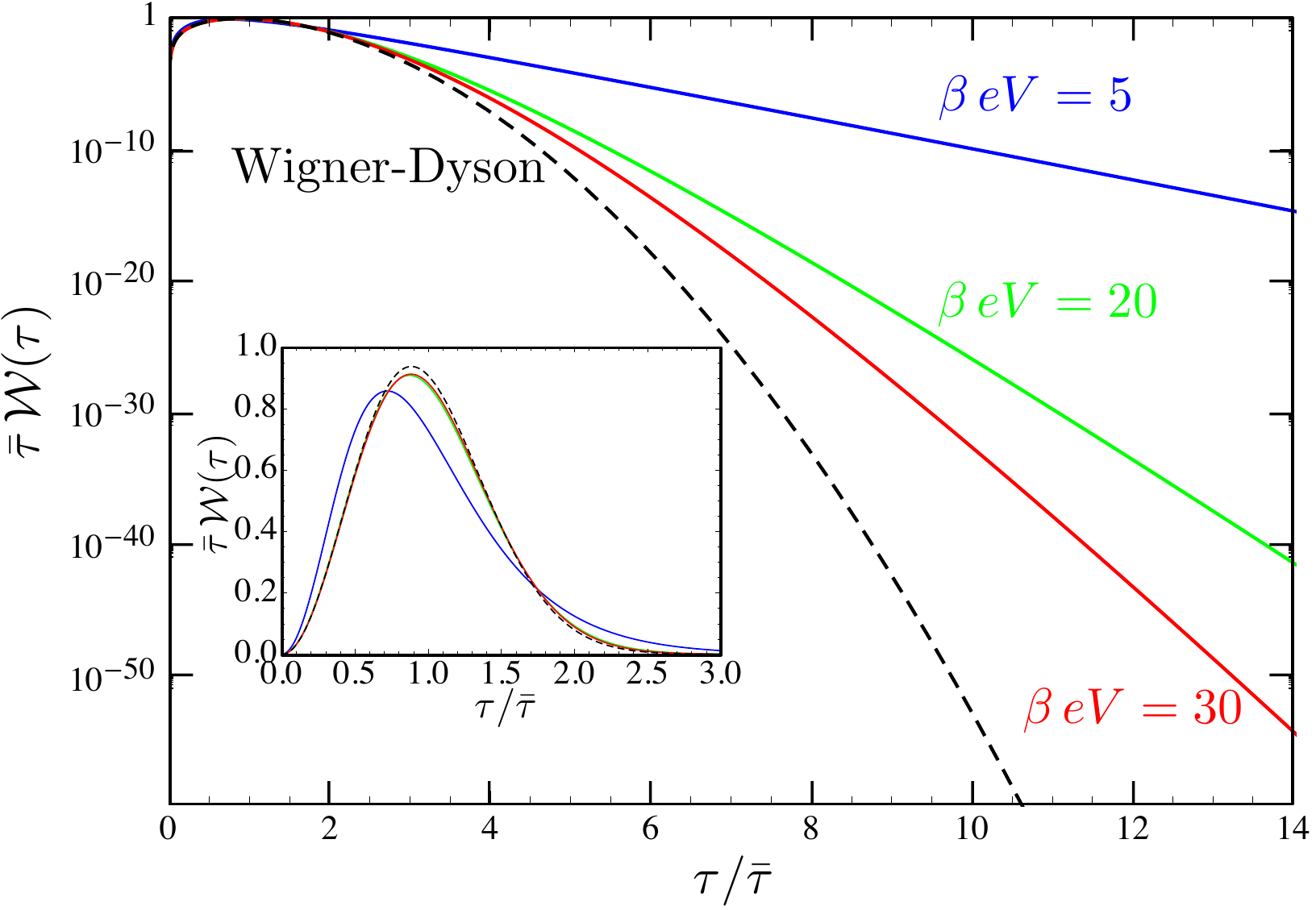}
  \caption{(Color online) WTD of a single-channel QPC with a finite temperature of the incoming electrons. The dashed line shows the Wigner-Dyson distribution, while colored lines correspond to finite temperatures ($\beta$ is the inverse temperature). The inset shows the same results on a linear scale. \label{fig:temp1}}
\end{figure}

To conclude this section, we return to the single-channel case, but now include a finite temperature of the incoming electrons. It is of particular interest to consider a fully open channel as the electronic temperature is increased. Figure \ref{fig:temp1} shows that the essential shape of the WTD, given by the Wigner-Dyson distribution, is robust against a non-zero electronic temperature on the order of the applied voltage bias. The tails of the WTD, on the other hand, change from a Gaussian decay at zero temperature to an exponential decay as the temperature approaches the applied voltage. We note that such a crossover has also been predicted for the emptiness formation probability in antiferromagnetic spin chains.\cite{Abanov2006} Finally, we mention that the small oscillations with period $\bar{\tau}$ observed for low transmissions disappear as the temperature is increased (not shown). \revision{Both effects can be understood by considering Eqs.~(\ref{eq:Qelements_finiteT}-\ref{eq:g}). The  finite temperature smears the Fermi distribution of the source electrode, allowing electrons with an energy greater than $eV$ to flow through the system. The temperature defines a new energy scale which makes the oscillations with period $\bar{\tau}$ disappear.}

\subsection{Fabry-P\'erot interferometer}

\begin{figure}
  \includegraphics[width=0.95\columnwidth]{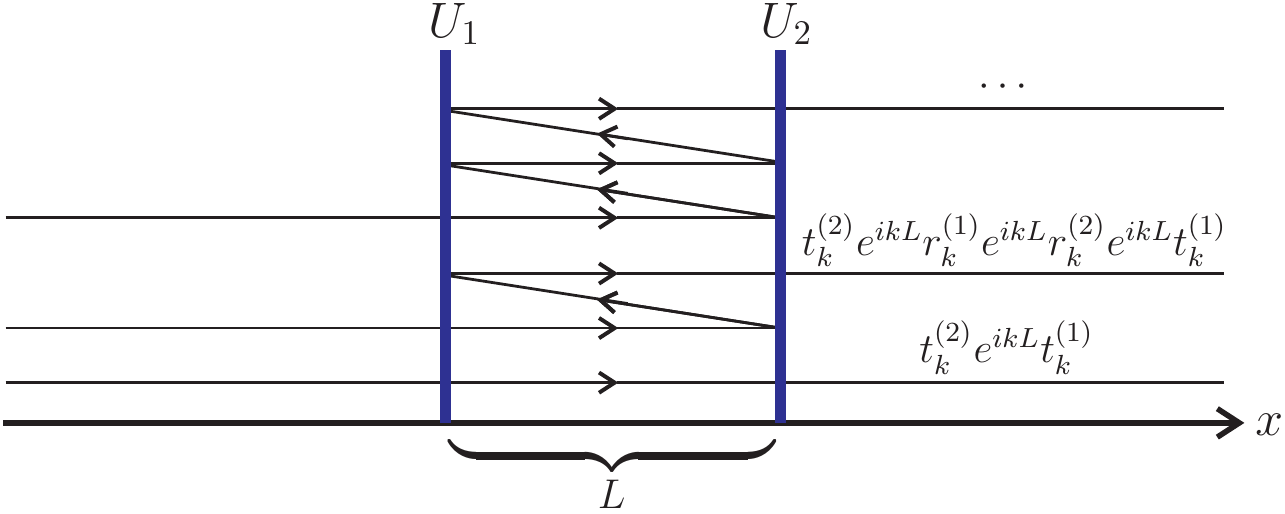}
    \caption{(Color online) Fabry-P\'erot interferometer. The interferometer consists of two delta-barriers of strength $U_j$, $j=1,2$, separated by the distance $L$. Transmitting trajectories with zero, two, and four internal reflections are indicated. The total transmission amplitude $t_k^{\mathrm{tot}}$ is the sum of the amplitudes of all transmitting trajectories.
  \label{fig:fp_setup}}
\end{figure}

As an example of a scatterer with energy-dependent transmission we consider a double-barrier structure consisting of delta-barriers of strengths $U_j$, $j=1,2$, separated by the distance $L$, see Fig.~\ref{fig:fp_setup}. The structure acts as a Fabry-P\'erot interferometer for electrons with a transmission probability that displays a series of peaks corresponding to resonant states of the interferometer.

The transmission and reflection amplitudes of the individual barriers are
\begin{equation}
t^{(j)}_k =\frac{1}{1+i\frac{mU_j}{\hbar^2 k}}, \,\,\, r^{(j)}_k =\frac{1}{1-i\frac{\hbar^2 k}{mU_j}}.
\end{equation}
As illustrated in Fig.~\ref{fig:fp_setup}, the total transmission amplitude of the double-barrier structure is the sum of the amplitudes of all transmitting trajectories
\begin{equation}
\begin{split}
t^{\mathrm{tot}}_k=t^{(2)}_ke^{ikL}t^{(1)}_k\sum_{n=0}^\infty (r^{(1)}_ke^{2ikL}r^{(2)}_k)^n=\frac{t^{(1)}_kt^{(2)}_ke^{ikL}}{1-r^{(1)}_kr^{(2)}_ke^{2ikL}}.
\end{split}
\end{equation}
The corresponding transmission probability reads
\begin{equation}
T^{\mathrm{tot}}_k=|t^{\mathrm{tot}}_k|^2=\frac{T^{(1)}_kT^{(2)}_k}{1+R^{(1)}_kR^{(2)}_k-2\sqrt{R^{(1)}_kR^{(2)}_k}\cos\theta_k}
\end{equation}
with $T^{(j)}_k=|t^{(j)}_k|^2$, $R^{(j)}_k=|r^{(j)}_k|^2$,  and $\theta_k=2kL+\arg r^{(1)}_k+\arg r^{(2)}_k$. For identical barriers, the transmission probability  reduces to
\begin{equation}
T^{\mathrm{tot}}_k=\frac{T^2}{1+R^2-2R\cos\theta_k},
\label{eq:FP-transmission}
\end{equation}
where $T = T^{(1)} = T^{(2)} = 1-R$ is the transmission of the individual barriers. We have assumed that the strengths of the delta barriers $U_1=U_2$ is of the order of $E_F/k_F$, such that $t_k^{(1)}=t_k^{(2)}$ varies slowly in the transport window and can be treated as $k$-independent. The total transmission is unity for values of $k$, where $\cos\theta_k=1$. These resonances are separated by the distance
\begin{equation}
\Delta k=\frac{\pi}{L}.
\end{equation}

We first consider a single resonance situated well inside the transport window. Specifically, we choose the Fermi energy and the applied voltage such that the transport window corresponds to the symmetric interval $[k_n-\Delta k/2, k_n+\Delta k/2]$ around a resonance at $k_n$. \revision{We note that it is possible to make the applied voltage larger than the width of the resonance, while still being much smaller than the Fermi energy, $\hbar v_F \Delta k< eV\ll E_F$.}

\begin{figure}
\includegraphics[width=0.92\columnwidth]{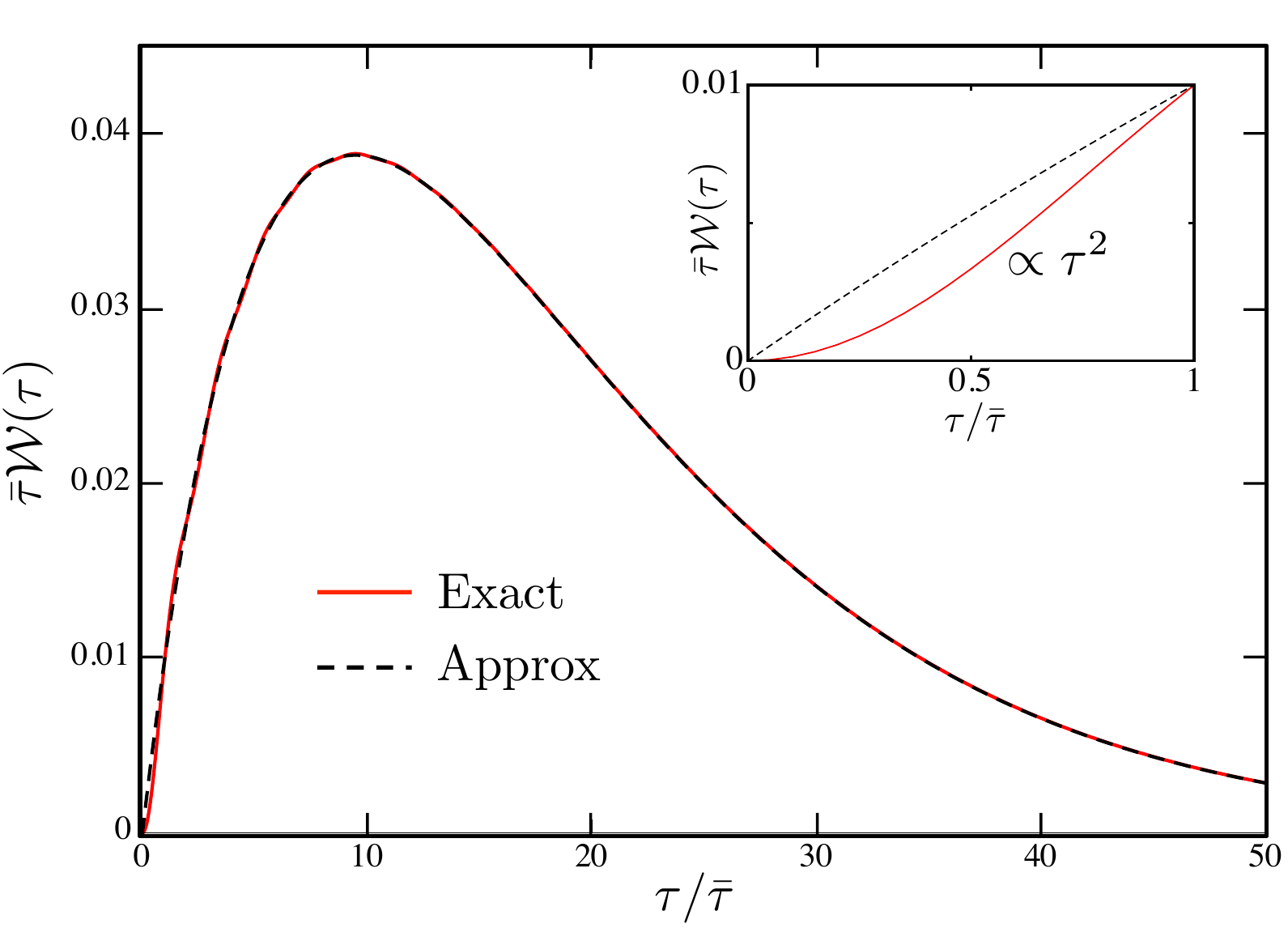}
  \caption{(Color online) WTD for a Fabry-P\'erot interferometer with a single resonance in the transport window. The transmissions of the individual barriers are $T^{(1)}= T^{(2)} = 0.1$.  The full line shows the exact results, while the dashed line indicates the approximation in Eq.~(\ref{eq:two-state-model}). The inset shows the short-time behavior, where the two curves clearly differ.}
\label{fig:resFP1}
\end{figure}

Figure~\ref{fig:resFP1} shows the results of our calculations. As we will now see, the WTD can be understood within the Breit-Wigner approximation. For small transmissions of the individual barriers, $T\ll1$, the total transmission probability is well-captured by the expression\cite{Lesovik2011}
\begin{equation}
T^{\mathrm{BW}}_k\simeq \sum_{n=0}^{\infty} \frac{(\gamma/2)^2}{(k-k_0-n\Delta k)^2+(\gamma/2)^2}
\label{eq:trans_BW}
\end{equation}
where
\begin{equation}
\gamma=\frac{1}{L}\frac{T}{\sqrt{1-T}}
\end{equation}
is the full width at half maximum and $k_0$ denotes the position of the resonance corresponding to $n = 0$. We interpret $L\gamma$ as the success probability of an electron to tunnel through the first barrier and we take $1/\bar{\tau}$ as the attempt frequency. Their product defines the rate
\begin{equation}
\Gamma=\frac{L\gamma}{\bar{\tau}}=\frac{eV}{h}\frac{T}{\sqrt{1-T}}
\label{eq:rate}
\end{equation}
at which electrons enter and leave the interferometer. From a simple rate equation calculation we then obtain an analytic expression for the WTD reading\cite{Brandes2008}
\begin{equation}
\W(\tau)=\Gamma^2\tau e^{-\Gamma\tau}.
\label{eq:two-state-model}
\end{equation}

As shown in Fig.~\ref{fig:resFP1}, the analytic expression shows surprisingly good agreement with the exact results over a large range of waiting times. Still, a closer inspection of the WTD reveals some deviations at short times $\tau\lesssim\bar{\tau}$, see lower inset in Fig.~\ref{fig:resFP1}. The simple rate model predicts a linear short-time behavior, $\W(\tau)\simeq \Gamma^2\tau$, whereas the exact results follow a quadratic dependence. Indeed, at short times, the WTD is determined by the waiting times between the in-coming electrons (given by the Wigner-Dyson distribution) rather than by the scatterer. This leads to the quadratic behavior at short times given in Eq.~(\ref{eq:wtd_short_times}). In contrast, the simple analytical model only describes the scatterer by the rate $\Gamma$ at which single electrons enter and leave the interferometer, giving rise to the linear short-time behavior.

It is instructive to compare our results for the waiting times with the Wigner-Smith delay time\cite{Wigner1955,Smith1960} of the Fabry-P\'{e}rot interferometer. We have calculated numerically the Wigner-Smith delay time at the center of a resonance and found that it corresponds to the dwell-time obtained from the rate equation description
\begin{equation}
\tau_{\mathrm{WS}}\simeq \tau_{\mathrm{dwell}}=1/\Gamma.
\end{equation}
This should be contrasted with the mean waiting time
\begin{equation}
\langle\tau\rangle=2/\Gamma,
\end{equation}
following from the rate equation description. The factor of two accounts for the fact that it takes an average time of $1/\Gamma$ to occupy the resonance after it has been emptied, followed by an average time of $1/\Gamma$ to empty it again.

\begin{figure}
  \includegraphics[width=0.92\columnwidth]{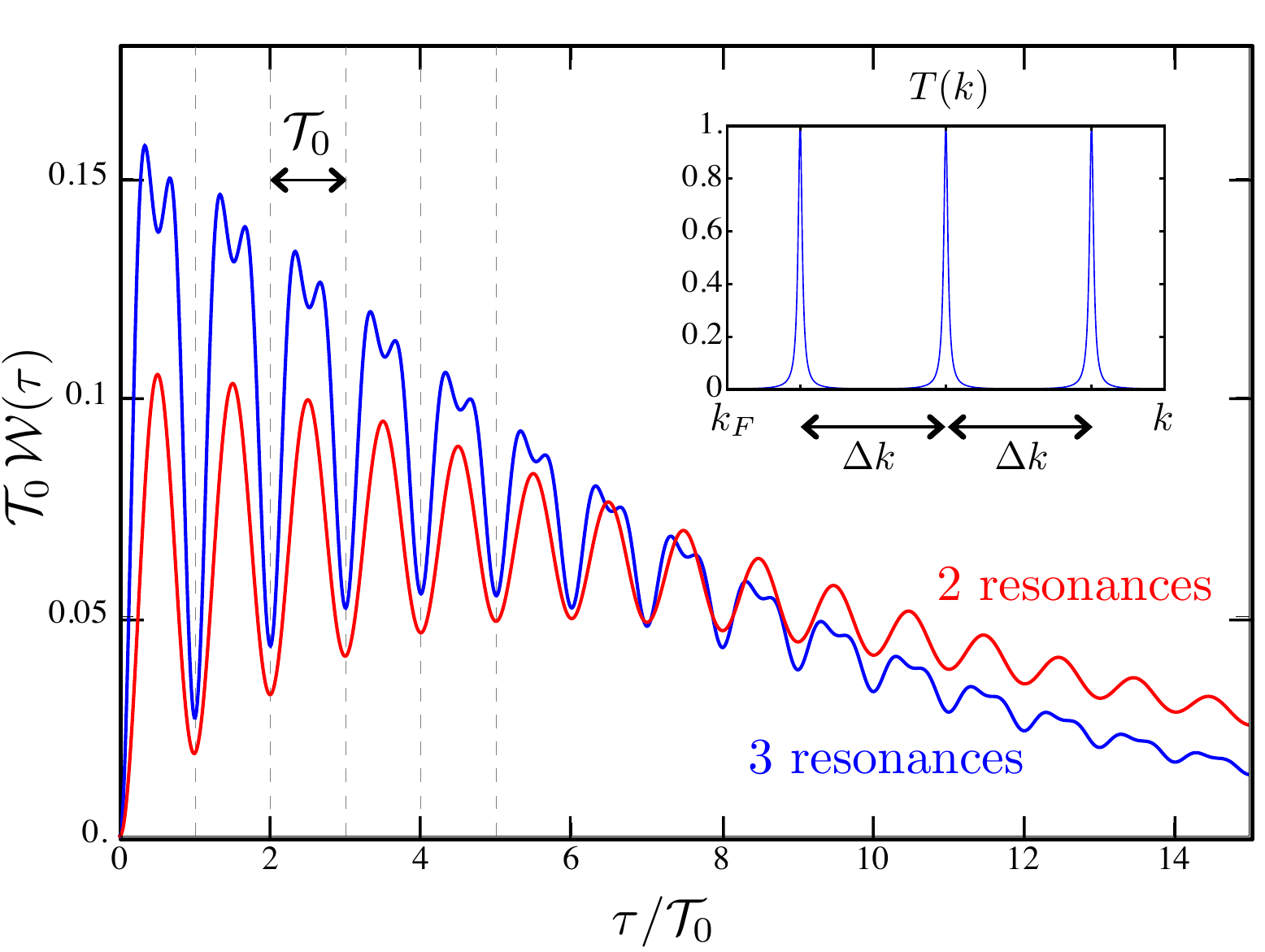}
  \caption{(Color online) WTD for a Fabry-P\'erot interferometer with two and three resonances inside the transport window. The transmission probabilities of the individual barriers are $T^{(1)}= T^{(2)} = 0.1$. The resonances are separated by $\Delta k=\pi/L$ as indicated in the inset showing the momentum-dependent transmission probability. With two resonances in the transport window, the WTD displays oscillations with period $\mathcal{T}_0=2\pi/(v_F\Delta k)$. With three resonances, additional small oscillations are seen with the period $\mathcal{T}_0/2$.}
\label{fig:resFP2}
\end{figure}

In Fig.~\ref{fig:resFP2} we increase the applied voltage so that two and three resonances are within the transport window, respectively. With two resonances in the transport window, we expect to see coherent oscillations in the WTD due to the interference between different energetic pathways. Indeed, denoting the energies of the two resonances as $E_1$ and $E_2$, respectively, with
\begin{equation}
E_2-E_1=\hbar v_F\Delta k,
\end{equation}
the single-particle interference pattern of the two energetic pathways becomes
\begin{equation}
\left|e^{-iE_1t/\hbar}+e^{-iE_2t/\hbar}\right|^2=2+2\cos\left(v_F\Delta kt\right),
\end{equation}
showing that the period of the oscillations should be
\begin{equation}
\mathcal{T}_0=\frac{2\pi}{v_F\Delta k}=\frac{2L}{v_F}.
\end{equation}
This period is clearly reflected in the WTD for the Fabry-P\'erot interferometer with two resonances in the transport window, see Fig.~\ref{fig:resFP2}. Extending the line of arguments to three equidistant resonances, we find
\begin{equation}
\begin{split}
&\left|e^{-iE_1t/h}+e^{-iE_2t/h}+e^{-iE_3t/h}\right|^2=\\
&3+4\cos\left(v_F\Delta kt\right)+2\cos\left(2v_F\Delta kt\right),
\end{split}
\end{equation}
showing that the oscillations with period $\mathcal{T}_0$ will be accompanied by smaller-amplitude oscillations at half the period. This behavior is clearly reflected in the WTD with three resonances shown in Fig.~\ref{fig:resFP2}. From Figs.~\ref{fig:resFP1} and~\ref{fig:resFP2} we conclude that the number of resonant levels contributing to the electronic transport can be inferred from the WTD.

\begin{figure}
  \includegraphics[width=0.94\columnwidth]{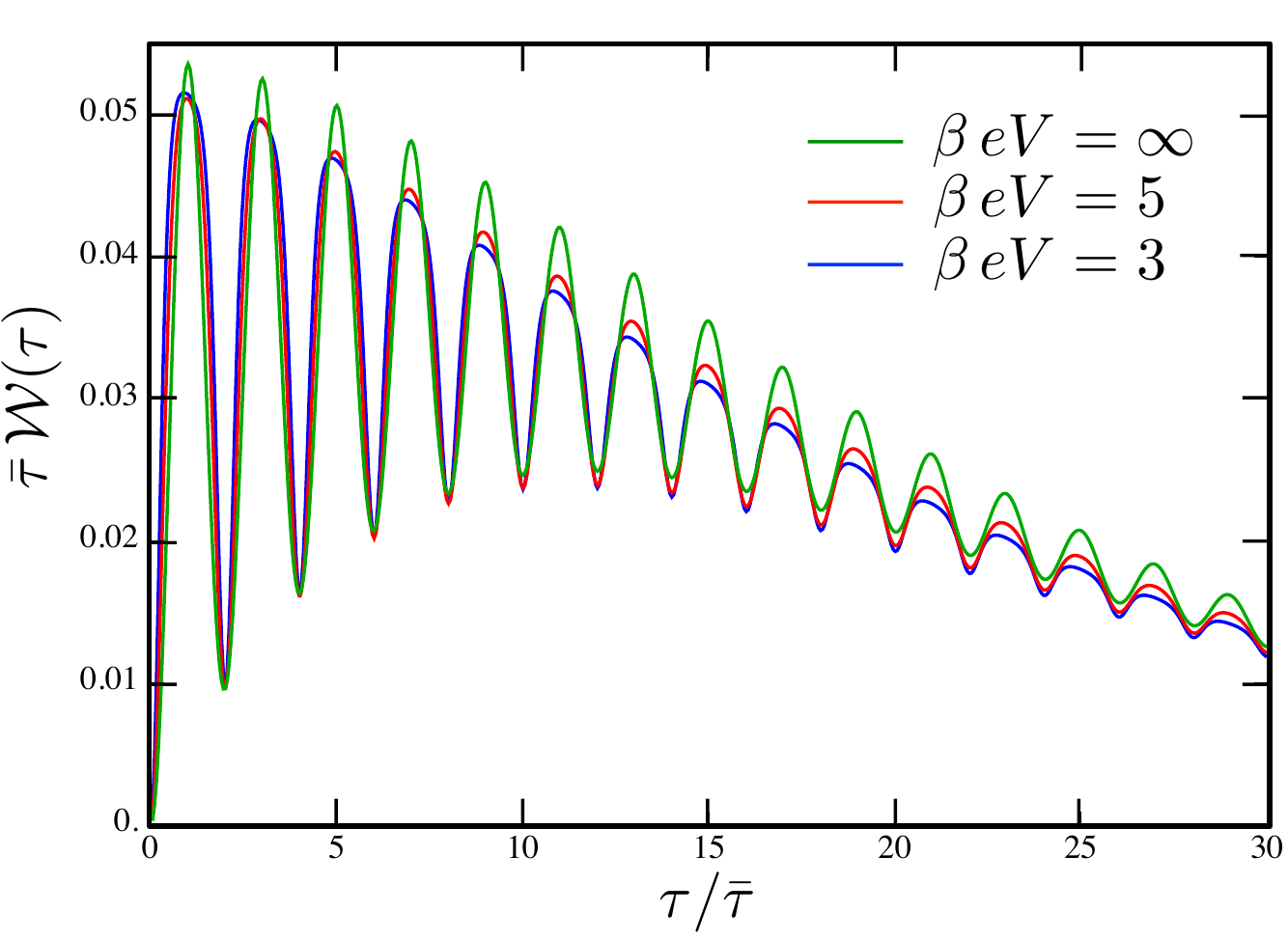}
  \caption{\revision{(Color online) WTD for a Fabry-P\'erot interferometer with two resonances inside the transport window and a finite electronic temperature. The transmission probabilities of the individual barriers are $T^{(1)}= T^{(2)} = 0.1$. The amplitude of the coherent oscillations is decreased as the electronic temperature of the source electrode is increased.}}
\label{fig:resFP2temp}
\end{figure}

\revision{In Fig.~\ref{fig:resFP2temp} we increase the temperature of the source electrode. With increasing temperature, the amplitude of the coherent oscillations decreases. We note that interference oscillations in WTDs have also been investigated for electronic transport in quantum dots with and without interactions.\cite{Brandes2008,Thomas2013,Thomas2014} In those cases, oscillations in the WTD were washed out by increasing the temperature of an external heat bath consisting of bosonic modes.}

\subsection{Mach-Zehnder interferometer}

\begin{figure}
\includegraphics[width=0.92\columnwidth]{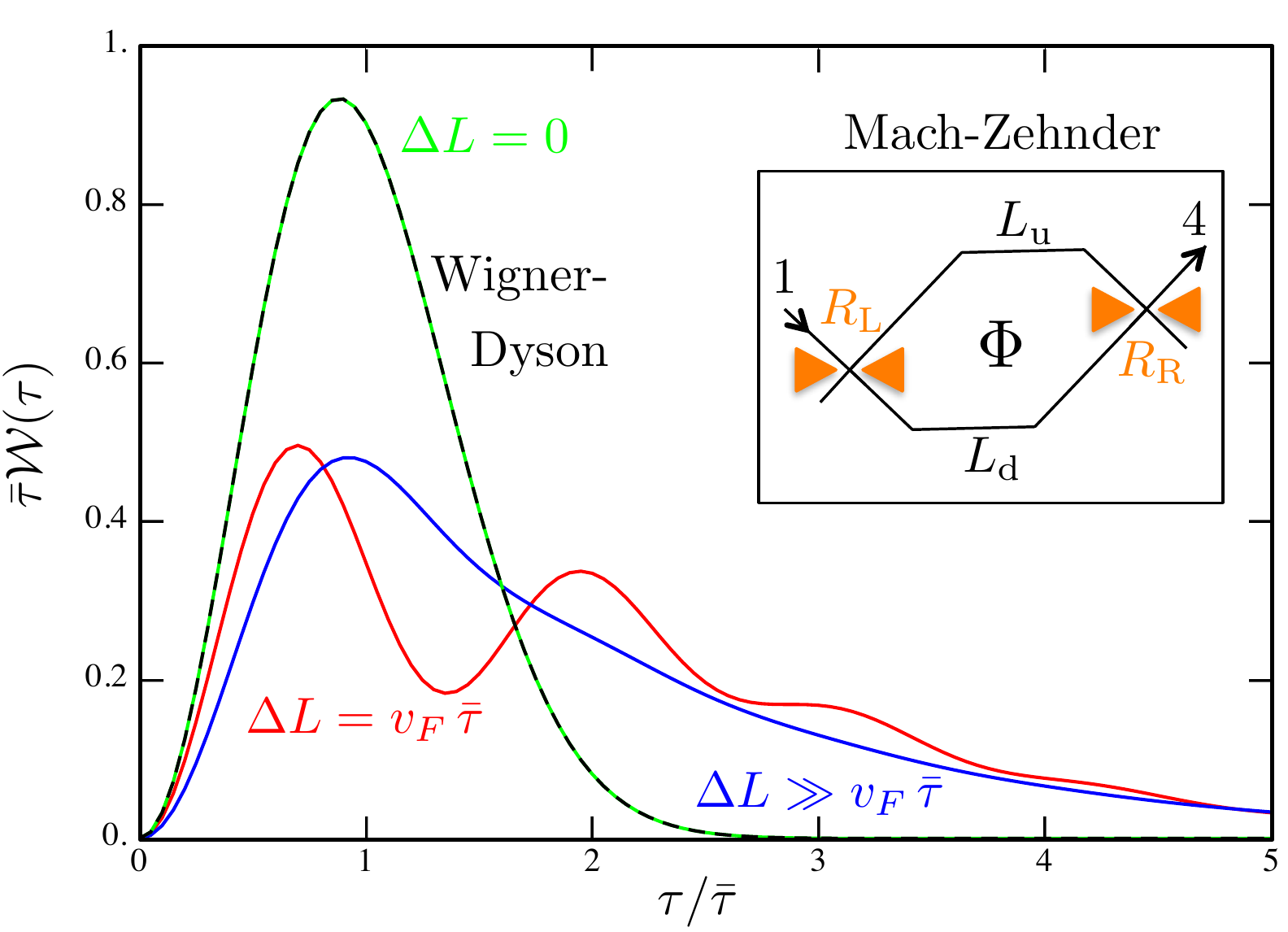}
\caption{(Color online) WTD for a Mach-Zehnder interferometer. The inset shows the interferometer made of a Corbino disk in the quantum Hall regime. Results are shown for half-transmitting QPCs, $R_\mathrm{R}=R_\mathrm{L} =T_\mathrm{R}=T_\mathrm{L}=1/2$, and three different path length differences $\Delta L$. The applied magnetic flux equals the flux quantum, $\Phi=\Phi_0$. For $\Delta L=0$, the WTD is well-captured by a Wigner-Dyson distribution. For $\Delta L \gg v_F\bar{\tau}$, the WTD corresponds to that of a QPC with transmission probability $R_\mathrm{R} R_\mathrm{L} + T_\mathrm{R} T_\mathrm{L}$ (not shown).
\label{fig:MZI}}
\end{figure}

As our last application, we consider the electronic Mach-Zehnder interferometer depicted in the inset of Fig.~\ref{fig:MZI}. The interferometer is made of a Corbino disk in the quantum Hall regime with electronic motion along edge states. Two QPCs act as electronic beam splitters.\cite{Ji2003,Neder2006,Roulleau2008,Litvin2008,Bieri2009,Roulleau2009}  The total transmission amplitude of the interferometer can be tuned either by changing the transmission probabilities of the individual QPCs,  the magnetic flux enclosed by the two arms, or the length of the arms using side-gate voltages. The Mach-Zehnder interferometer is of special interest as it makes it possible to determine the coherence time of the incoming single-electron states.\cite{Haack2011, Haack2013} The coherence time is defined as the time span during which single-particle interferences are observable at the output of the interferometer. For a dc-biased source, the coherence time is given by $\bar{\tau}$. Below, we investigate the influence of single-particle interferences on the WTD of a Mach-Zehnder interferometer.

To evaluate the total transmission amplitude we assume that the QPCs have energy-independent transmissions. The momentum-dependent transmission amplitude $t_{k}^{41}$ to go from input 1 to output 4 then reads
\begin{eqnarray}
t_{k}^{41} &=& -\sqrt{R_\mathrm{L} R_\mathrm{R}} e^{i (kL_\mathrm{u} + \theta_\mathrm{u})} + \sqrt{T_\mathrm{L} T_\mathrm{R}} e^{i(k L_\mathrm{d} + \theta_\mathrm{d})}.
\end{eqnarray}
The coefficients $R_\mathrm{L,R}$ and $T_\mathrm{L,R}$ are the reflection and transmission probabilities of the left and right QPCs, respectively. The lengths of the upper and lower arms are denoted as $L_\mathrm{u,d}$, and $\theta_\mathrm{u,d}$ are the magnetic phases such that
\begin{equation}
\theta_\mathrm{u} - \theta_\mathrm{d} = 2\pi \Phi/\Phi_0,
\end{equation}
where $\Phi$ is the enclosed magnetic flux and $\Phi_0 = h/e$ is the magnetic flux quantum. With this transmission amplitude, the matrix elements in Eq.~(\ref{eq:Qelements}) become
\begin{equation}
\label{eq:Q_MZI}
\begin{split}
[\bold{Q}_\tau]_{m,n} &= \frac{\sin(\kappa v_F  \tau (n-m)/2)}{\pi(n-m)} e^{-i\kappa v_F  \tau (n-m)/2}\times  \\
& \left[ R_\mathrm{R} R_\mathrm{L} e^{i\kappa\Delta L (n-m) /2} + T_\mathrm{R} T_\mathrm{L} e^{-i\kappa\Delta L(n-m) /2}\right.\\
&+ \left.2 \sqrt{R_\mathrm{R} R_\mathrm{L}T_\mathrm{R} T_\mathrm{L}} \cos\left(2\pi \Phi/\Phi_0 + \kappa \Delta L(n+m)/2\right) \right],
\end{split}
\nonumber
\end{equation}
where $\Delta L=L_\mathrm{u}-L_\mathrm{d}$ is the path length difference.

In Fig.~\ref{fig:MZI} we show results for the WTD with varying path length differences. With no path length difference, $\Delta L = 0$, the transmission probability becomes $k$-independent and simplifies to
\begin{equation}
T^{41} = R_\mathrm{R} R_\mathrm{L} + T_\mathrm{R} T_\mathrm{L} + 2 \sqrt{R_\mathrm{R} R_\mathrm{L}T_\mathrm{R} T_\mathrm{L}} \cos\left(2\pi \Phi/\Phi_0\right).
\label{eq:MZI_small_asym}
\nonumber
\end{equation}
Here, the two first terms are ``classical'' contributions corresponding to electrons that are transmitted from terminal 1 to terminal 4 by either being reflected at both QPCs or being transmitted through both. By contrast, the third term is due to single-particle interferences which are sensitive to the applied magnetic field. Interestingly, by choosing half-transmitting QPCs, $R_\mathrm{R}=R_\mathrm{L} =T_\mathrm{R}=T_\mathrm{L}=1/2$, the interferometer becomes fully transmitting if the magnetic flux is tuned to an integer number of magnetic flux quanta, $\Phi=n\Phi_0$,  and fully reflecting if $\Phi=(n+1/2)\Phi_0$, where $n$ is an integer. In Fig.~\ref{fig:MZI}, we show results for the fully transmitting case and find that the WTD again is well-captured by a Wigner-Dyson distribution.

In the other limiting case, where the path length difference is very large, $\Delta L \gg v_F\bar{\tau}$, single-particle interferences are suppressed. In this case, the WTD corresponds to that of a QPC with a transmission probability given by the classical contributions $R_\mathrm{R} R_\mathrm{L} + T_\mathrm{R} T_\mathrm{L}$. Finally, in the intermediate regime, $\Delta L \simeq v_F\bar{\tau}$, the interference part contributes and the WTD exhibits oscillations whose period depends on the phase of the transmission amplitude, which is now momentum-dependent. The oscillations in the WTD, which are due to the single-particle interferences, are clearly seen in Fig.~\ref{fig:MZI}. As such, the WTD constitutes an alternative approach to observe single-particle interferences in addition to measurements of the average current and the shot noise.

\section{Conclusions and outlook}
\label{sec:conclusions}

We have presented a detailed description of our theory of electron waiting times in mesoscopic transport and extended earlier results for single-channel conductors to setups with several (possibly spin-degenerate) conduction channels as well as finite electronic temperatures. As a specific application, we have analyzed in detail the distribution of electron waiting times in a quantum point contact (QPC). We have shown that the waiting time distribution (WTD) for a QPC can be understood using a mapping between one-dimensional free fermions and the level spacing statistics of random matrices. The WTD displays a crossover from Wigner-Dyson statistics at full transmission to Poisson statistics close to pinch-off. The suppression of the WTD at short times due to the Pauli principle is lifted as the number of conduction channels is increased.

As examples of scatterers with energy-dependent transmissions, we have investigated the WTDs of a Fabry-P\'{e}rot interferometer and a Mach-Zehnder interferometer. For the Fabry-P\'{e}rot interferometer, oscillations in the WTD reflect the number of resonant levels contributing to the electronic transport. With a single resonance well inside the bias window, the WTD can be understood using a simple rate equation description with rates obtained from a Breit-Wigner approximation of the transmission probability. With several resonances in the transport window, the WTD displays quantum mechanical interference oscillations. For the Mach-Zehnder interferometer, the WTD shows signatures of single-particle interferences which can be controlled by adjusting the path length difference and the applied magnetic field.

Our work leaves a number of interesting tasks and questions to be addressed in the future. Our theory implicitly assumes a single-electron detector that is only sensitive to electrons above the Fermi level. Experimentally, progress is currently being made towards the realization of such a detector.\cite{Thalineau2014} A theoretical description of a single-electron detector with a built-in energy filter is also under development. Another interesting question concerns correlations between subsequent waiting times. For a renewal process, the waiting times are uncorrelated.\cite{Cox1962} However, there are several indications that electronic transport in mesoscopic conductors cannot be described as a renewal process.\cite{Albert2012,Dasenbrook2014} To investigate this question in further detail, a theory of joint distributions of waiting times is needed. Finally, it may be possible to link the WTDs to specific correlations functions describing the incoming many-body state.\cite{Verstraete2010,Hubener2013}

\section*{Acknowledgements}

The paper is dedicated to the memory of Markus B\"uttiker. We thank him for his continuous support and for many inspiring discussions. We also acknowledge several useful discussions with the late Oriol Bohigas. Finally, we thank David Dasenbrook, Patrick P.~Hofer, Bj\"{o}rn Sothmann, and Konrad H.~Thomas for useful comments on the manuscript. The work was supported by the NCCR QSIT, SNSF, and the Alexander von Humboldt Foundation in the framework of the Alexander von Humboldt Professorship.

\appendix

\section{Analogy with random matrix theory}
\label{AppenB}

Here we explain the mapping between random matrices and one-dimensional fermions. We note that this mapping does not correspond to a phenomenological description of a mesoscopic device with a random Hamiltonian.

We consider the canonical Gaussian ensembles of random matrices introduced by Wigner\cite{Haake2001,Mehta2004} to understand certain universal properties of complex quantum systems like nuclei. The constraints on the allowed Hamiltonians depend on the symmetries of the problem, for instance time-reversal symmetry or spin, leading to different universality classes labeled by the symmetry index $\beta$. The Gaussian orthogonal ensemble (time-reversal invariance, but no spin) has $\beta=1$ and corresponds to real symmetric matrices. The Gaussian unitary ensemble (breaking of time-reversal symmetry, and no spin) has $\beta=2$ and corresponds to hermitian matrices.

The matrix elements $H_{ij}$ of the Hamiltonian $\mathbf{H}$ of rank $N$ are chosen randomly from the distribution
\begin{equation}
  P(\{H_{ij}\})=\mathcal N_{N,\beta} \exp[-\beta\, \textrm{Tr} \{\mathbf{H} ^2\}/v^2],
\end{equation}
where $\mathcal N_{N,\beta}$ is a normalization constant and $v$ a typical energy scale. After diagonalization of the Hamiltonian, the joint probability of the eigenvalues $\{E_i\}$ becomes
\begin{equation}
  P(\{E_i\})=\mathcal C_{N,\beta}\, e^{-\frac{\beta}{v^2} \sum_j E_j^2} \prod_{i<j} |E_i-E_j|^\beta,
\end{equation}
where $\mathcal C_{N,\beta}$ again is a normalization constant. Importantly, this probability distribution can be interpreted as the  square modulus of a fermionic wave function in one dimension, if ones replaces the eigenenergies $\{E_i\}$ by fictitious quantum particles at positions $\{x_i\}$. The corresponding many-body wave function reads
\begin{equation}
  \Psi(\{x_i\})=\sqrt{\mathcal C_{N,\beta}} e^{-\frac{\beta}{2 v^2}\sum_j x_j^2} \prod_{i<j} |x_i-x_j|^{\beta/2}\textrm{sgn}(x_i-x_j),
    \nonumber
\end{equation}
where the sign function $\textrm{sgn}(x_i-x_j)$ expresses the anti-symmetric nature of fermionic many-body wave functions. The terms $|x_i-x_j|^{\beta/2}$ give rise to repulsion between particles, whereas the exponentials come from a harmonic confining potential, which can be made arbitrarily weak in the large-$N$ limit with an appropriate rescaling of the $x_i$'s.

One can show that $\Psi(\{x_i\})$ in fact is the ground state of the interacting Calogero-Sutherland Hamiltonian\cite{Calogero1969,Sutherland1971}
\begin{equation}
  H=-\frac{1}{2}\sum_{i=1}^N \frac{\partial^2}{\partial x_i^2}+\frac{\beta}{2}\left(\frac{\beta}{2}-1\right)\sum_{i>j} \frac{1}{(x_i-x_j)^2}
\end{equation}
with an inverse square interaction. For $\beta=2$, the interactions vanish and all correlations are due only to the fermionic nature of the fictitious particles. This shows us that the correlation functions of free fermions are formally equivalent to those obtained from random matrix theory. We note that this mapping also makes it possible to compute the WTDs of interacting fermions in two special cases, namely $\beta=1$ (attractive interactions) and $\beta=4$ (repulsive interactions), corresponding to the Gaussian orthogonal ensemble and the Gaussian symplectic ensemble, respectively.

\section{Renewal theory}
\label{AppenA}

Here we briefly illustrate how various statistical quantities can be obtained from the WTD of a renewal process. For a renewal process, subsequent waiting times are statistically independent and essentially all information is encoded in the WTD.\cite{Cox1962} This makes it possible, for instance, to infer the probability $P(n,t)$ of observing $n$ events during a long time span of duration $t$. This probability can be written as
\begin{equation}
P(n,t) = \int_0^\infty\!\! d\tau_1\cdots  \int_0^\infty\!\! d\tau_n \W(\tau_1)\cdots\W(\tau_n)\delta(t-\sum_{i=1}^n\tau_i),
\nonumber
\end{equation}
having assumed that an event occurs at the beginning and the end of the interval, but only one of them is counted. This assumption is not important for the long-time limit that we consider here.

Corresponding to $P(n,t)$ we define the moment generating function
\begin{equation}
M(\chi,t)=\sum_n P(n,t)e^{in\chi}.
\end{equation}
Simple algebra allows us to write the moment generating function as
\begin{equation}
M(\chi,t)=\frac{1}{2\pi i}\int_{-i\infty}^{i\infty}dz \frac{e^{z t}}{1-e^{i\chi+\log\widetilde{\W}(z)}},
\label{eq:MGF}
\end{equation}
where
\begin{equation}
\widetilde{\W}(z)=\int_0^\infty d\tau \W(\tau)e^{-z\tau}
\end{equation}
is the Laplace transform of the WTD and we have used a Fourier representation of the delta function in the expression for $P(n,t)$. We note that $\log\widetilde{\W}(z)$ generates the cumulants of the waiting time by differentiation with respect to $-z$ at $z=0$,
\begin{equation}
\llangle \tau^m\rrangle = (-1)^m \frac{d^m}{dz^m} \log\widetilde{\W}(z)|_{z=0}.
\label{eq:WTDcumulants}
\end{equation}

At long times, the moment generating function becomes exponential in time, $M(\chi,t)\propto e^{z_0(\chi) t}$, with a rate determined by the pole $z_0(\chi)$ of the integrand in Eq.\ (\ref{eq:MGF}) closest to $z=0$. The cumulant generating function for $n$ is defined as $S(\chi,t)=\log M(\chi,t)$ which becomes linear in time, $S(\chi,t)\rightarrow z_0(\chi) t$. We can therefore identify $z_0(\chi)$ as the cumulant generating function of the charge current, which yields the current cumulants by differentiation with respect to $i\chi$ at $\chi=0$,
\begin{equation}
\langle\!\langle I^m\rangle\!\rangle = \frac{d^m}{d(i \chi)^m} z_0(\chi)|_{\chi=0}.
\label{eq:Curcumulants}
\end{equation}
To find the cumulant generating function of the current we need to solve the equation
\begin{equation}
i\chi+\log\widetilde{\W}(z)=0
\label{eq:WTD2CGF}
\end{equation}
for the pole $z=z_0(\chi)$ closest to $z=0$ with $z_0(0)=0$. In general, this is a difficult task. However, by consecutively differentiating Eq.\ (\ref{eq:WTD2CGF}) evaluated at $z=0$ we can establish the following relations between the cumulants of the waiting time and the cumulants of the current\cite{Albert2011}
\begin{equation}
\label{eq:cumulants}
   \begin{split}
   \frac{\llangle I^2\rrangle}{\llangle I\rrangle}& =\frac{\llangle \tau^2\rrangle}{\llangle \tau\rrangle^2},\\
   \frac{\llangle I^3\rrangle}{\llangle I\rrangle}&=3\frac{\llangle \tau^2\rrangle^2}{\llangle \tau\rrangle^4}-\frac{\llangle \tau^3\rrangle}{\llangle \tau\rrangle^3},\\
   \frac{\llangle I^4\rrangle}{\llangle I\rrangle}&= 15 \frac{\llangle \tau^2\rrangle^3}{\llangle \tau\rrangle^6} - 10 \frac{\llangle \tau^2\rrangle \llangle \tau^3\rrangle}{\llangle \tau\rrangle^5}+\frac{\llangle \tau^4\rrangle}{\llangle \tau\rrangle^4}.
   \end{split}
 \end{equation}
These relations illustrate how the current cumulants for a renewal process are directly related to the cumulants of the waiting time. For a renewal process, the $g^{(2)}$-function can also be directly obtained from the WTD, see for example Refs.~\onlinecite{Carmichael1989, Emary2012}.


\begin{thebibliography}{65}%
\makeatletter
\providecommand \@ifxundefined [1]{%
 \@ifx{#1\undefined}
}%
\providecommand \@ifnum [1]{%
 \ifnum #1\expandafter \@firstoftwo
 \else \expandafter \@secondoftwo
 \fi
}%
\providecommand \@ifx [1]{%
 \ifx #1\expandafter \@firstoftwo
 \else \expandafter \@secondoftwo
 \fi
}%
\providecommand \natexlab [1]{#1}%
\providecommand \enquote  [1]{``#1''}%
\providecommand \bibnamefont  [1]{#1}%
\providecommand \bibfnamefont [1]{#1}%
\providecommand \citenamefont [1]{#1}%
\providecommand \href@noop [0]{\@secondoftwo}%
\providecommand \href [0]{\begingroup \@sanitize@url \@href}%
\providecommand \@href[1]{\@@startlink{#1}\@@href}%
\providecommand \@@href[1]{\endgroup#1\@@endlink}%
\providecommand \@sanitize@url [0]{\catcode `\\12\catcode `\$12\catcode
  `\&12\catcode `\#12\catcode `\^12\catcode `\_12\catcode `\%12\relax}%
\providecommand \@@startlink[1]{}%
\providecommand \@@endlink[0]{}%
\providecommand \url  [0]{\begingroup\@sanitize@url \@url }%
\providecommand \@url [1]{\endgroup\@href {#1}{\urlprefix }}%
\providecommand \urlprefix  [0]{URL }%
\providecommand \Eprint [0]{\href }%
\@ifxundefined \urlstyle {%
  \providecommand \doi  [0]{\begingroup \@sanitize@url \@doi}%
  \providecommand \@doi [1]{\endgroup \@@startlink {\doibase
  #1}doi:\discretionary {}{}{}#1\@@endlink }%
}{%
  \providecommand \doi  [0]{doi:\discretionary{}{}{}\begingroup
  \urlstyle{rm}\Url }%
}%
\providecommand \doibase [0]{http://dx.doi.org/}%
\providecommand \Doi [0]{\begingroup \@sanitize@url \@Doi }%
\providecommand \@Doi  [1]{\endgroup\@@startlink{\doibase#1}\@@Doi}%
\providecommand \@@Doi [1]{#1\@@endlink}%
\providecommand \selectlanguage [0]{\@gobble}%
\providecommand \bibinfo  [0]{\@secondoftwo}%
\providecommand \bibfield  [0]{\@secondoftwo}%
\providecommand \translation [1]{[#1]}%
\providecommand \BibitemOpen [0]{}%
\providecommand \bibitemStop [0]{}%
\providecommand \bibitemNoStop [0]{.\EOS\space}%
\providecommand \EOS [0]{\spacefactor3000\relax}%
\providecommand \BibitemShut  [1]{\csname bibitem#1\endcsname}%
\bibitem [{\citenamefont {Blanter}\ and\ \citenamefont
  {B\"{u}ttiker}(2000)}]{Blanter2000}%
  \BibitemOpen
  \bibfield  {author} {\bibinfo {author} {\bibfnamefont {Ya.~M.}\ \bibnamefont
  {Blanter}}\ and\ \bibinfo {author} {\bibfnamefont {M.}~\bibnamefont
  {B\"{u}ttiker}},\ }\Doi {10.1016/S0370-1573(99)00123-4} {\bibfield  {journal}
  {\bibinfo  {journal} {Phys. Rep.},\ }\textbf {\bibinfo {volume} {336}},\
  \bibinfo {pages} {1} (\bibinfo {year} {2000})}\BibitemShut {NoStop}%
\bibitem [{\citenamefont {Levitov}\ and\ \citenamefont
  {Lesovik}(1993)}]{Levitov1993}%
  \BibitemOpen
  \bibfield  {author} {\bibinfo {author} {\bibfnamefont {L.~S.}\ \bibnamefont
  {Levitov}}\ and\ \bibinfo {author} {\bibfnamefont {G.~B.}\ \bibnamefont
  {Lesovik}},\ }\href@noop {} {\bibfield  {journal} {\bibinfo  {journal} {JETP
  Lett.},\ }\textbf {\bibinfo {volume} {58}},\ \bibinfo {pages} {230} (\bibinfo
  {year} {1993})}\BibitemShut {NoStop}%
\bibitem [{\citenamefont {Levitov}\ \emph {et~al.}(1996)\citenamefont
  {Levitov}, \citenamefont {Lee},\ and\ \citenamefont {Lesovik}}]{Levitov1996}%
  \BibitemOpen
  \bibfield  {author} {\bibinfo {author} {\bibfnamefont {L.~S.}\ \bibnamefont
  {Levitov}}, \bibinfo {author} {\bibfnamefont {H.}~\bibnamefont {Lee}}, \ and\
  \bibinfo {author} {\bibfnamefont {G.~B.}\ \bibnamefont {Lesovik}},\ }\Doi
  {10.1063/1.531672} {\bibfield  {journal} {\bibinfo  {journal} {J. Math.
  Phys.},\ }\textbf {\bibinfo {volume} {37}},\ \bibinfo {pages} {4845}
  (\bibinfo {year} {1996})}\BibitemShut {NoStop}%
\bibitem [{\citenamefont {Nazarov}(2003)}]{Nazarov2003}%
  \BibitemOpen
  \bibinfo {editor} {\bibfnamefont {Yu.~V.}\ \bibnamefont {Nazarov}},\ ed.,\
  \href@noop {} {\emph {\bibinfo {title} {Quantum Noise in Mesoscopic
  Physics}}}\ (\bibinfo  {publisher} {Kluwer, Dordrecht},\ \bibinfo {year}
  {2003})\BibitemShut {NoStop}%
\bibitem [{\citenamefont {Reulet}\ \emph {et~al.}(2003)\citenamefont {Reulet},
  \citenamefont {Senzier},\ and\ \citenamefont {Prober}}]{Reulet2003}%
  \BibitemOpen
  \bibfield  {author} {\bibinfo {author} {\bibfnamefont {B.}~\bibnamefont
  {Reulet}}, \bibinfo {author} {\bibfnamefont {J.}~\bibnamefont {Senzier}}, \
  and\ \bibinfo {author} {\bibfnamefont {D.~E.}\ \bibnamefont {Prober}},\ }\Doi
  {10.1103/PhysRevLett.91.196601} {\bibfield  {journal} {\bibinfo  {journal}
  {Phys. Rev. Lett.},\ }\textbf {\bibinfo {volume} {91}},\ \bibinfo {pages}
  {196601} (\bibinfo {year} {2003})}\BibitemShut {NoStop}%
\bibitem [{\citenamefont {Bylander}\ \emph {et~al.}(2005)\citenamefont
  {Bylander}, \citenamefont {Duty},\ and\ \citenamefont
  {Delsing}}]{Bylander2005}%
  \BibitemOpen
  \bibfield  {author} {\bibinfo {author} {\bibfnamefont {J.}~\bibnamefont
  {Bylander}}, \bibinfo {author} {\bibfnamefont {T.}~\bibnamefont {Duty}}, \
  and\ \bibinfo {author} {\bibfnamefont {P.}~\bibnamefont {Delsing}},\ }\Doi
  {10.1038/nature03375} {\bibfield  {journal} {\bibinfo  {journal} {Nature},\
  }\textbf {\bibinfo {volume} {434}},\ \bibinfo {pages} {361} (\bibinfo {year}
  {2005})}\BibitemShut {NoStop}%
\bibitem [{\citenamefont {Bomze}\ \emph {et~al.}(2005)\citenamefont {Bomze},
  \citenamefont {Gershon}, \citenamefont {Shovkun}, \citenamefont {Levitov},\
  and\ \citenamefont {Reznikov}}]{Bomze2005}%
  \BibitemOpen
  \bibfield  {author} {\bibinfo {author} {\bibfnamefont {Yu.}~\bibnamefont
  {Bomze}}, \bibinfo {author} {\bibfnamefont {G.}~\bibnamefont {Gershon}},
  \bibinfo {author} {\bibfnamefont {D.}~\bibnamefont {Shovkun}}, \bibinfo
  {author} {\bibfnamefont {L.~S.}\ \bibnamefont {Levitov}}, \ and\ \bibinfo
  {author} {\bibfnamefont {M.}~\bibnamefont {Reznikov}},\ }\Doi
  {10.1103/PhysRevLett.95.176601} {\bibfield  {journal} {\bibinfo  {journal}
  {Phys. Rev. Lett.},\ }\textbf {\bibinfo {volume} {95}},\ \bibinfo {pages}
  {176601} (\bibinfo {year} {2005})}\BibitemShut {NoStop}%
\bibitem [{\citenamefont {Gustavsson}\ \emph {et~al.}(2006)\citenamefont
  {Gustavsson}, \citenamefont {Leturcq}, \citenamefont {Simovic}, \citenamefont
  {Schleser}, \citenamefont {Ihn}, \citenamefont {Studerus}, \citenamefont
  {Ensslin}, \citenamefont {Driscoll},\ and\ \citenamefont
  {Gossard}}]{Gustavsson2006}%
  \BibitemOpen
  \bibfield  {author} {\bibinfo {author} {\bibfnamefont {S.}~\bibnamefont
  {Gustavsson}}, \bibinfo {author} {\bibfnamefont {R.}~\bibnamefont {Leturcq}},
  \bibinfo {author} {\bibfnamefont {B.}~\bibnamefont {Simovic}}, \bibinfo
  {author} {\bibfnamefont {R.}~\bibnamefont {Schleser}}, \bibinfo {author}
  {\bibfnamefont {T.}~\bibnamefont {Ihn}}, \bibinfo {author} {\bibfnamefont
  {P.}~\bibnamefont {Studerus}}, \bibinfo {author} {\bibfnamefont
  {K.}~\bibnamefont {Ensslin}}, \bibinfo {author} {\bibfnamefont {D.~C.}\
  \bibnamefont {Driscoll}}, \ and\ \bibinfo {author} {\bibfnamefont {A.~C.}\
  \bibnamefont {Gossard}},\ }\Doi {10.1103/PhysRevLett.96.076605} {\bibfield
  {journal} {\bibinfo  {journal} {Phys. Rev. Lett.},\ }\textbf {\bibinfo
  {volume} {96}},\ \bibinfo {pages} {076605} (\bibinfo {year}
  {2006})}\BibitemShut {NoStop}%
\bibitem [{\citenamefont {Fujisawa}\ \emph {et~al.}(2006)\citenamefont
  {Fujisawa}, \citenamefont {Hayashi}, \citenamefont {Tomita},\ and\
  \citenamefont {Hirayama}}]{Fujisawa2006}%
  \BibitemOpen
  \bibfield  {author} {\bibinfo {author} {\bibfnamefont {T.}~\bibnamefont
  {Fujisawa}}, \bibinfo {author} {\bibfnamefont {T.}~\bibnamefont {Hayashi}},
  \bibinfo {author} {\bibfnamefont {R.}~\bibnamefont {Tomita}}, \ and\ \bibinfo
  {author} {\bibfnamefont {Y.}~\bibnamefont {Hirayama}},\ }\Doi
  {10.1126/science.1126788} {\bibfield  {journal} {\bibinfo  {journal}
  {Science},\ }\textbf {\bibinfo {volume} {312}},\ \bibinfo {pages} {1634}
  (\bibinfo {year} {2006})}\BibitemShut {NoStop}%
\bibitem [{\citenamefont {Fricke}\ \emph {et~al.}(2007)\citenamefont {Fricke},
  \citenamefont {Hohls}, \citenamefont {Wegscheider},\ and\ \citenamefont
  {Haug}}]{Fricke2007}%
  \BibitemOpen
  \bibfield  {author} {\bibinfo {author} {\bibfnamefont {C.}~\bibnamefont
  {Fricke}}, \bibinfo {author} {\bibfnamefont {F.}~\bibnamefont {Hohls}},
  \bibinfo {author} {\bibfnamefont {W.}~\bibnamefont {Wegscheider}}, \ and\
  \bibinfo {author} {\bibfnamefont {R.~J.}\ \bibnamefont {Haug}},\ }\Doi
  {10.1103/PhysRevB.76.155307} {\bibfield  {journal} {\bibinfo  {journal}
  {Phys. Rev. B},\ }\textbf {\bibinfo {volume} {76}},\ \bibinfo {pages}
  {155307} (\bibinfo {year} {2007})}\BibitemShut {NoStop}%
\bibitem [{\citenamefont {Sukhorukov}\ \emph {et~al.}(2007)\citenamefont
  {Sukhorukov}, \citenamefont {Jordan}, \citenamefont {Gustavsson},
  \citenamefont {Leturcq}, \citenamefont {Ihn},\ and\ \citenamefont
  {Ensslin}}]{Sukhorukov2007}%
  \BibitemOpen
  \bibfield  {author} {\bibinfo {author} {\bibfnamefont {E.~V.}\ \bibnamefont
  {Sukhorukov}}, \bibinfo {author} {\bibfnamefont {A.~N.}\ \bibnamefont
  {Jordan}}, \bibinfo {author} {\bibfnamefont {S.}~\bibnamefont {Gustavsson}},
  \bibinfo {author} {\bibfnamefont {R.}~\bibnamefont {Leturcq}}, \bibinfo
  {author} {\bibfnamefont {T.}~\bibnamefont {Ihn}}, \ and\ \bibinfo {author}
  {\bibfnamefont {K.}~\bibnamefont {Ensslin}},\ }\Doi {10.1038/nphys564}
  {\bibfield  {journal} {\bibinfo  {journal} {Nature Physics},\ }\textbf
  {\bibinfo {volume} {3}},\ \bibinfo {pages} {243} (\bibinfo {year}
  {2007})}\BibitemShut {NoStop}%
\bibitem [{\citenamefont {Timofeev}\ \emph {et~al.}(2007)\citenamefont
  {Timofeev}, \citenamefont {Meschke}, \citenamefont {Peltonen}, \citenamefont
  {Heikkila},\ and\ \citenamefont {Pekola}}]{Timofeev2007}%
  \BibitemOpen
  \bibfield  {author} {\bibinfo {author} {\bibfnamefont {A.~V.}\ \bibnamefont
  {Timofeev}}, \bibinfo {author} {\bibfnamefont {M.}~\bibnamefont {Meschke}},
  \bibinfo {author} {\bibfnamefont {J.~T.}\ \bibnamefont {Peltonen}}, \bibinfo
  {author} {\bibfnamefont {T.~T.}\ \bibnamefont {Heikkila}}, \ and\ \bibinfo
  {author} {\bibfnamefont {J.~P.}\ \bibnamefont {Pekola}},\ }\Doi
  {10.1103/PhysRevLett.98.207001} {\bibfield  {journal} {\bibinfo  {journal}
  {Phys Rev. Lett.},\ }\textbf {\bibinfo {volume} {98}},\ \bibinfo {pages}
  {207001} (\bibinfo {year} {2007})}\BibitemShut {NoStop}%
\bibitem [{\citenamefont {Gershon}\ \emph {et~al.}(2008)\citenamefont
  {Gershon}, \citenamefont {Bomze}, \citenamefont {Sukhorukov},\ and\
  \citenamefont {Reznikov}}]{Gershon2008}%
  \BibitemOpen
  \bibfield  {author} {\bibinfo {author} {\bibfnamefont {G.}~\bibnamefont
  {Gershon}}, \bibinfo {author} {\bibfnamefont {Y.}~\bibnamefont {Bomze}},
  \bibinfo {author} {\bibfnamefont {E.~V.}\ \bibnamefont {Sukhorukov}}, \ and\
  \bibinfo {author} {\bibfnamefont {M.}~\bibnamefont {Reznikov}},\ }\Doi
  {10.1103/PhysRevLett.101.016803} {\bibfield  {journal} {\bibinfo  {journal}
  {Phys. Rev. Lett.},\ }\textbf {\bibinfo {volume} {101}},\ \bibinfo {pages}
  {016803} (\bibinfo {year} {2008})}\BibitemShut {NoStop}%
\bibitem [{\citenamefont {Flindt}\ \emph {et~al.}(2009)\citenamefont {Flindt},
  \citenamefont {Hohls}, \citenamefont {Novotn\'{y}}, \citenamefont
  {Neto\v{c}n\'{y}}, \citenamefont {Brandes},\ and\ \citenamefont
  {Haug}}]{Flindt2009}%
  \BibitemOpen
  \bibfield  {author} {\bibinfo {author} {\bibfnamefont {C.}~\bibnamefont
  {Flindt}, \bibfnamefont {C.~Fricke}}, \bibinfo {author} {\bibfnamefont
  {F.}~\bibnamefont {Hohls}}, \bibinfo {author} {\bibfnamefont
  {T.}~\bibnamefont {Novotn\'{y}}}, \bibinfo {author} {\bibfnamefont
  {K.}~\bibnamefont {Neto\v{c}n\'{y}}}, \bibinfo {author} {\bibfnamefont
  {T.}~\bibnamefont {Brandes}}, \ and\ \bibinfo {author} {\bibfnamefont
  {R.~J.}\ \bibnamefont {Haug}},\ }\Doi {10.1073/pnas.0901002106} {\bibfield
  {journal} {\bibinfo  {journal} {Proc. Natl. Acad. Sci. USA},\ }\textbf
  {\bibinfo {volume} {106}},\ \bibinfo {pages} {10119} (\bibinfo {year}
  {2009})}\BibitemShut {NoStop}%
\bibitem [{\citenamefont {Gabelli}\ and\ \citenamefont
  {Reulet}(2009)}]{Gabelli2009}%
  \BibitemOpen
  \bibfield  {author} {\bibinfo {author} {\bibfnamefont {J.}~\bibnamefont
  {Gabelli}}\ and\ \bibinfo {author} {\bibfnamefont {B.}~\bibnamefont
  {Reulet}},\ }\href@noop {} {\bibfield  {journal} {\bibinfo  {journal} {Phys.
  Rev. B},\ }\textbf {\bibinfo {volume} {80}},\ \bibinfo {pages} {161203}
  (\bibinfo {year} {2009})}\BibitemShut {NoStop}%
\bibitem [{\citenamefont {Le~Masne}\ \emph {et~al.}(2009)\citenamefont
  {Le~Masne}, \citenamefont {Pothier}, \citenamefont {Birge}, \citenamefont
  {Urbina},\ and\ \citenamefont {Esteve}}]{Masne2009}%
  \BibitemOpen
  \bibfield  {author} {\bibinfo {author} {\bibfnamefont {Q.}~\bibnamefont
  {Le~Masne}}, \bibinfo {author} {\bibfnamefont {H.}~\bibnamefont {Pothier}},
  \bibinfo {author} {\bibfnamefont {N.~O.}\ \bibnamefont {Birge}}, \bibinfo
  {author} {\bibfnamefont {C.}~\bibnamefont {Urbina}}, \ and\ \bibinfo {author}
  {\bibfnamefont {D.}~\bibnamefont {Esteve}},\ }\Doi
  {10.1103/PhysRevLett.102.067002} {\bibfield  {journal} {\bibinfo  {journal}
  {Phys. Rev. Lett.},\ }\textbf {\bibinfo {volume} {102}},\ \bibinfo {pages}
  {067002} (\bibinfo {year} {2009})}\BibitemShut {NoStop}%
\bibitem [{\citenamefont {Fricke}\ \emph
  {et~al.}(2010){\natexlab{a}}\citenamefont {Fricke}, \citenamefont {Hohls},
  \citenamefont {Flindt},\ and\ \citenamefont {Haug}}]{Fricke2010}%
  \BibitemOpen
  \bibfield  {author} {\bibinfo {author} {\bibfnamefont {C.}~\bibnamefont
  {Fricke}}, \bibinfo {author} {\bibfnamefont {F.}~\bibnamefont {Hohls}},
  \bibinfo {author} {\bibfnamefont {C.}~\bibnamefont {Flindt}}, \ and\ \bibinfo
  {author} {\bibfnamefont {R.~J.}\ \bibnamefont {Haug}},\ }\Doi
  {http://dx.doi.org/10.1016/j.physe.2009.11.112} {\bibfield  {journal}
  {\bibinfo  {journal} {Physica E},\ }\textbf {\bibinfo {volume} {42}},\
  \bibinfo {pages} {848} (\bibinfo {year} {2010}{\natexlab{a}})}\BibitemShut
  {NoStop}%
\bibitem [{\citenamefont {Fricke}\ \emph
  {et~al.}(2010){\natexlab{b}}\citenamefont {Fricke}, \citenamefont {Hohls},
  \citenamefont {Sethubalasubramanian}, \citenamefont {Fricke},\ and\
  \citenamefont {Haug}}]{Fricke2010b}%
  \BibitemOpen
  \bibfield  {author} {\bibinfo {author} {\bibfnamefont {C.}~\bibnamefont
  {Fricke}}, \bibinfo {author} {\bibfnamefont {F.}~\bibnamefont {Hohls}},
  \bibinfo {author} {\bibfnamefont {N.}~\bibnamefont {Sethubalasubramanian}},
  \bibinfo {author} {\bibfnamefont {L.}~\bibnamefont {Fricke}}, \ and\ \bibinfo
  {author} {\bibfnamefont {R.~J.}\ \bibnamefont {Haug}},\ }\Doi
  {dx.doi.org/10.1063/1.3430000} {\bibfield  {journal} {\bibinfo  {journal}
  {Appl. Phys. Lett.},\ }\textbf {\bibinfo {volume} {96}},\ \bibinfo {pages}
  {202103} (\bibinfo {year} {2010}{\natexlab{b}})}\BibitemShut {NoStop}%
\bibitem [{\citenamefont {Ubbelohde}\ \emph {et~al.}(2012)\citenamefont
  {Ubbelohde}, \citenamefont {Fricke}, \citenamefont {Flindt}, \citenamefont
  {Hohls},\ and\ \citenamefont {Haug}}]{Ubbelohde2012}%
  \BibitemOpen
  \bibfield  {author} {\bibinfo {author} {\bibfnamefont {N.}~\bibnamefont
  {Ubbelohde}}, \bibinfo {author} {\bibfnamefont {C.}~\bibnamefont {Fricke}},
  \bibinfo {author} {\bibfnamefont {C.}~\bibnamefont {Flindt}}, \bibinfo
  {author} {\bibfnamefont {F.}~\bibnamefont {Hohls}}, \ and\ \bibinfo {author}
  {\bibfnamefont {R.~J.}\ \bibnamefont {Haug}},\ }\Doi {10.1038/ncomms1620}
  {\bibfield  {journal} {\bibinfo  {journal} {Nat. Commun.},\ }\textbf
  {\bibinfo {volume} {3}},\ \bibinfo {pages} {612} (\bibinfo {year}
  {2012})}\BibitemShut {NoStop}%
\bibitem [{\citenamefont {Maisi}\ \emph {et~al.}(2014)\citenamefont {Maisi},
  \citenamefont {Kambly}, \citenamefont {Flindt},\ and\ \citenamefont
  {Pekola}}]{Maisi2014}%
  \BibitemOpen
  \bibfield  {author} {\bibinfo {author} {\bibfnamefont {V.~F.}\ \bibnamefont
  {Maisi}}, \bibinfo {author} {\bibfnamefont {D.}~\bibnamefont {Kambly}},
  \bibinfo {author} {\bibfnamefont {C.}~\bibnamefont {Flindt}}, \ and\ \bibinfo
  {author} {\bibfnamefont {J.~P.}\ \bibnamefont {Pekola}},\ }\Doi
  {10.1103/PhysRevLett.112.036801} {\bibfield  {journal} {\bibinfo  {journal}
  {Phys. Rev. Lett.},\ }\textbf {\bibinfo {volume} {112}},\ \bibinfo {pages}
  {036801} (\bibinfo {year} {2014})}\BibitemShut {NoStop}%
\bibitem [{\citenamefont {Brandes}(2008)}]{Brandes2008}%
  \BibitemOpen
  \bibfield  {author} {\bibinfo {author} {\bibfnamefont {T.}~\bibnamefont
  {Brandes}},\ }\Doi {10.1002/andp.200810306} {\bibfield  {journal} {\bibinfo
  {journal} {Ann. Phys. (Berlin)},\ }\textbf {\bibinfo {volume} {17}},\
  \bibinfo {pages} {477} (\bibinfo {year} {2008})}\BibitemShut {NoStop}%
\bibitem [{\citenamefont {Welack}\ \emph {et~al.}(2008)\citenamefont {Welack},
  \citenamefont {Esposito}, \citenamefont {Harbola},\ and\ \citenamefont
  {Mukamel}}]{Welack2008}%
  \BibitemOpen
  \bibfield  {author} {\bibinfo {author} {\bibfnamefont {S.}~\bibnamefont
  {Welack}}, \bibinfo {author} {\bibfnamefont {M.}~\bibnamefont {Esposito}},
  \bibinfo {author} {\bibfnamefont {U.}~\bibnamefont {Harbola}}, \ and\
  \bibinfo {author} {\bibfnamefont {S.}~\bibnamefont {Mukamel}},\ }\Doi
  {10.1103/PhysRevB.77.195315} {\bibfield  {journal} {\bibinfo  {journal}
  {Phys. Rev. B},\ }\textbf {\bibinfo {volume} {77}},\ \bibinfo {pages}
  {195315} (\bibinfo {year} {2008})}\BibitemShut {NoStop}%
\bibitem [{\citenamefont {Welack}\ \emph {et~al.}(2009)\citenamefont {Welack},
  \citenamefont {Mukamel},\ and\ \citenamefont {Yan}}]{Welack2009}%
  \BibitemOpen
  \bibfield  {author} {\bibinfo {author} {\bibfnamefont {S.}~\bibnamefont
  {Welack}}, \bibinfo {author} {\bibfnamefont {S.}~\bibnamefont {Mukamel}}, \
  and\ \bibinfo {author} {\bibfnamefont {Y.}~\bibnamefont {Yan}},\ }\Doi
  {10.1209/0295-5075/85/57008} {\bibfield  {journal} {\bibinfo  {journal}
  {Europhys. Lett.},\ }\textbf {\bibinfo {volume} {85}},\ \bibinfo {pages}
  {57008} (\bibinfo {year} {2009})}\BibitemShut {NoStop}%
\bibitem [{\citenamefont {Albert}\ \emph {et~al.}(2011)\citenamefont {Albert},
  \citenamefont {Flindt},\ and\ \citenamefont {B\"{u}ttiker}}]{Albert2011}%
  \BibitemOpen
  \bibfield  {author} {\bibinfo {author} {\bibfnamefont {M.}~\bibnamefont
  {Albert}}, \bibinfo {author} {\bibfnamefont {C.}~\bibnamefont {Flindt}}, \
  and\ \bibinfo {author} {\bibfnamefont {M.}~\bibnamefont {B\"{u}ttiker}},\
  }\Doi {10.1103/PhysRevLett.107.086805} {\bibfield  {journal} {\bibinfo
  {journal} {Phys. Rev. Lett.},\ }\textbf {\bibinfo {volume} {107}},\ \bibinfo
  {pages} {086805} (\bibinfo {year} {2011})}\BibitemShut {NoStop}%
\bibitem [{\citenamefont {Albert}\ \emph {et~al.}(2012)\citenamefont {Albert},
  \citenamefont {Haack}, \citenamefont {Flindt},\ and\ \citenamefont
  {B\"{u}ttiker}}]{Albert2012}%
  \BibitemOpen
  \bibfield  {author} {\bibinfo {author} {\bibfnamefont {M.}~\bibnamefont
  {Albert}}, \bibinfo {author} {\bibfnamefont {G.}~\bibnamefont {Haack}},
  \bibinfo {author} {\bibfnamefont {C.}~\bibnamefont {Flindt}}, \ and\ \bibinfo
  {author} {\bibfnamefont {M.}~\bibnamefont {B\"{u}ttiker}},\ }\Doi
  {10.1103/PhysRevLett.108.186806} {\bibfield  {journal} {\bibinfo  {journal}
  {Phys. Rev. Lett.},\ }\textbf {\bibinfo {volume} {108}},\ \bibinfo {pages}
  {186806} (\bibinfo {year} {2012})}\BibitemShut {NoStop}%
\bibitem [{\citenamefont {Thomas}\ and\ \citenamefont
  {Flindt}(2013)}]{Thomas2013}%
  \BibitemOpen
  \bibfield  {author} {\bibinfo {author} {\bibfnamefont {K.~H.}\ \bibnamefont
  {Thomas}}\ and\ \bibinfo {author} {\bibfnamefont {C.}~\bibnamefont
  {Flindt}},\ }\Doi {10.1103/PhysRevB.87.121405} {\bibfield  {journal}
  {\bibinfo  {journal} {Phys. Rev. B},\ }\textbf {\bibinfo {volume} {87}},\
  \bibinfo {pages} {121405(R)} (\bibinfo {year} {2013})}\BibitemShut {NoStop}%
\bibitem [{\citenamefont {Rajabi}\ \emph {et~al.}(2013)\citenamefont {Rajabi},
  \citenamefont {P\"oltl},\ and\ \citenamefont {Governale}}]{Rajabi2013}%
  \BibitemOpen
  \bibfield  {author} {\bibinfo {author} {\bibfnamefont {L.}~\bibnamefont
  {Rajabi}}, \bibinfo {author} {\bibfnamefont {C.}~\bibnamefont {P\"oltl}}, \
  and\ \bibinfo {author} {\bibfnamefont {M.}~\bibnamefont {Governale}},\ }\Doi
  {10.1103/PhysRevLett.111.067002} {\bibfield  {journal} {\bibinfo  {journal}
  {Phys. Rev. Lett.},\ }\textbf {\bibinfo {volume} {111}},\ \bibinfo {pages}
  {067002} (\bibinfo {year} {2013})}\BibitemShut {NoStop}%
\bibitem [{\citenamefont {Dasenbrook}\ \emph {et~al.}(2014)\citenamefont
  {Dasenbrook}, \citenamefont {Flindt},\ and\ \citenamefont
  {B\"uttiker}}]{Dasenbrook2014}%
  \BibitemOpen
  \bibfield  {author} {\bibinfo {author} {\bibfnamefont {D.}~\bibnamefont
  {Dasenbrook}}, \bibinfo {author} {\bibfnamefont {C.}~\bibnamefont {Flindt}},
  \ and\ \bibinfo {author} {\bibfnamefont {M.}~\bibnamefont {B\"uttiker}},\
  }\Doi {10.1103/PhysRevLett.112.146801} {\bibfield  {journal} {\bibinfo
  {journal} {Phys. Rev. Lett.},\ }\textbf {\bibinfo {volume} {112}},\ \bibinfo
  {pages} {146801} (\bibinfo {year} {2014})}\BibitemShut {NoStop}%
\bibitem [{\citenamefont {Thomas}\ and\ \citenamefont
  {Flindt}(2014)}]{Thomas2014}%
  \BibitemOpen
  \bibfield  {author} {\bibinfo {author} {\bibfnamefont {K.~H.}\ \bibnamefont
  {Thomas}}\ and\ \bibinfo {author} {\bibfnamefont {C.}~\bibnamefont
  {Flindt}},\ }\Doi {10.1103/PhysRevB.89.245420} {\bibfield  {journal}
  {\bibinfo  {journal} {Phys. Rev. B},\ }\textbf {\bibinfo {volume} {89}},\
  \bibinfo {pages} {245420} (\bibinfo {year} {2014})}\BibitemShut {NoStop}%
\bibitem [{\citenamefont {Tang}\ \emph {et~al.}(2014)\citenamefont {Tang},
  \citenamefont {Xu},\ and\ \citenamefont {Wang}}]{Wang2014}%
  \BibitemOpen
  \bibfield  {author} {\bibinfo {author} {\bibfnamefont {G.-M.}\ \bibnamefont
  {Tang}}, \bibinfo {author} {\bibfnamefont {F.}~\bibnamefont {Xu}}, \ and\
  \bibinfo {author} {\bibfnamefont {J.}~\bibnamefont {Wang}},\ }\Doi
  {10.1103/PhysRevB.89.205310} {\bibfield  {journal} {\bibinfo  {journal}
  {Phys. Rev. B},\ }\textbf {\bibinfo {volume} {89}},\ \bibinfo {pages}
  {205310} (\bibinfo {year} {2014})}\BibitemShut {NoStop}%
\bibitem [{\citenamefont {Albert}\ and\ \citenamefont
  {Devillard}(2014)}]{Albert2014}%
  \BibitemOpen
  \bibfield  {author} {\bibinfo {author} {\bibfnamefont {M.}~\bibnamefont
  {Albert}}\ and\ \bibinfo {author} {\bibfnamefont {P.}~\bibnamefont
  {Devillard}},\ }\Doi {10.1103/PhysRevB.90.035431} {\bibfield  {journal}
  {\bibinfo  {journal} {Phys. Rev. B},\ }\textbf {\bibinfo {volume} {90}},\
  \bibinfo {pages} {035431} (\bibinfo {year} {2014})}\BibitemShut {NoStop}%
\bibitem [{\citenamefont {Sothmann}()}]{Sothmann2014}%
  \BibitemOpen
  \bibfield  {author} {\bibinfo {author} {\bibfnamefont {B.}~\bibnamefont
  {Sothmann}},\ }\Doi {10.1103/PhysRevB.90.155315} {\bibfield  {journal}
  {\bibinfo  {journal} {Phys. Rev. B},\ }\textbf {\bibinfo {volume} {90}},\
  \bibinfo {pages} {155315} (\bibinfo {year} {2014})}\BibitemShut {NoStop}%
\bibitem [{\citenamefont {Cox}(1962)}]{Cox1962}%
  \BibitemOpen
\bibfield  {journal} {  }\bibfield  {author} {\bibinfo {author} {\bibfnamefont
  {D.~R.}\ \bibnamefont {Cox}},\ }\href@noop {} {\emph {\bibinfo {title}
  {Renewal Theory}}}\ (\bibinfo  {publisher} {Methuen \& Co.},\ \bibinfo {year}
  {1962})\BibitemShut {NoStop}%
\bibitem [{\citenamefont {Vyas}\ and\ \citenamefont {Singh}(1988)}]{Vyas1988}%
  \BibitemOpen
  \bibfield  {author} {\bibinfo {author} {\bibfnamefont {R.}~\bibnamefont
  {Vyas}}\ and\ \bibinfo {author} {\bibfnamefont {S.}~\bibnamefont {Singh}},\
  }\Doi {10.1103/PhysRevA.38.2423} {\bibfield  {journal} {\bibinfo  {journal}
  {Phys. Rev. A},\ }\textbf {\bibinfo {volume} {38}},\ \bibinfo {pages} {2423}
  (\bibinfo {year} {1988})}\BibitemShut {NoStop}%
\bibitem [{\citenamefont {Carmichael}\ \emph {et~al.}(1989)\citenamefont
  {Carmichael}, \citenamefont {Singh}, \citenamefont {Vyas},\ and\
  \citenamefont {Rice}}]{Carmichael1989}%
  \BibitemOpen
  \bibfield  {author} {\bibinfo {author} {\bibfnamefont {H.~J.}\ \bibnamefont
  {Carmichael}}, \bibinfo {author} {\bibfnamefont {S.}~\bibnamefont {Singh}},
  \bibinfo {author} {\bibfnamefont {R.}~\bibnamefont {Vyas}}, \ and\ \bibinfo
  {author} {\bibfnamefont {P.~R.}\ \bibnamefont {Rice}},\ }\Doi
  {10.1103/PhysRevA.39.1200} {\bibfield  {journal} {\bibinfo  {journal} {Phys.
  Rev. A},\ }\textbf {\bibinfo {volume} {39}},\ \bibinfo {pages} {1200}
  (\bibinfo {year} {1989})}\BibitemShut {NoStop}%
\bibitem [{\citenamefont {Moskalets}(2011)}]{Moskalets2011}%
  \BibitemOpen
  \bibfield  {author} {\bibinfo {author} {\bibfnamefont {M.}~\bibnamefont
  {Moskalets}},\ }\href@noop {} {\emph {\bibinfo {title} {Scattering Matrix
  Approach to Non-Stationary Quantum Transport}}}\ (\bibinfo  {publisher}
  {Imperial College Press, London},\ \bibinfo {year} {2011})\BibitemShut
  {NoStop}%
\bibitem [{\citenamefont {F\`eve}\ \emph {et~al.}(2007)\citenamefont {F\`eve},
  \citenamefont {Mah\'e}, \citenamefont {Berroir}, \citenamefont {Kontos},
  \citenamefont {Pla\c{c}ais}, \citenamefont {Glattli}, \citenamefont
  {Cavanna}, \citenamefont {Etienne},\ and\ \citenamefont {Jin}}]{Feve2007}%
  \BibitemOpen
  \bibfield  {author} {\bibinfo {author} {\bibfnamefont {G.}~\bibnamefont
  {F\`eve}}, \bibinfo {author} {\bibfnamefont {A.}~\bibnamefont {Mah\'e}},
  \bibinfo {author} {\bibfnamefont {J.-M.}\ \bibnamefont {Berroir}}, \bibinfo
  {author} {\bibfnamefont {T.}~\bibnamefont {Kontos}}, \bibinfo {author}
  {\bibfnamefont {B.}~\bibnamefont {Pla\c{c}ais}}, \bibinfo {author}
  {\bibfnamefont {D.~C.}\ \bibnamefont {Glattli}}, \bibinfo {author}
  {\bibfnamefont {A.}~\bibnamefont {Cavanna}}, \bibinfo {author} {\bibfnamefont
  {B.}~\bibnamefont {Etienne}}, \ and\ \bibinfo {author} {\bibfnamefont
  {Y.}~\bibnamefont {Jin}},\ }\Doi {10.1126/science.1141243} {\bibfield
  {journal} {\bibinfo  {journal} {Science},\ }\textbf {\bibinfo {volume}
  {316}},\ \bibinfo {pages} {1169} (\bibinfo {year} {2007})}\BibitemShut
  {NoStop}%
\bibitem [{\citenamefont {Bocquillon}\ \emph {et~al.}(2013)\citenamefont
  {Bocquillon}, \citenamefont {Freulon}, \citenamefont {Berroir}, \citenamefont
  {Degiovanni}, \citenamefont {Pla\c{c}ais}, \citenamefont {Cavanna},
  \citenamefont {Jin},\ and\ \citenamefont {F\`eve}}]{Bocquillon2013}%
  \BibitemOpen
  \bibfield  {author} {\bibinfo {author} {\bibfnamefont {E.}~\bibnamefont
  {Bocquillon}}, \bibinfo {author} {\bibfnamefont {V.}~\bibnamefont {Freulon}},
  \bibinfo {author} {\bibfnamefont {J.-M.}\ \bibnamefont {Berroir}}, \bibinfo
  {author} {\bibfnamefont {P.}~\bibnamefont {Degiovanni}}, \bibinfo {author}
  {\bibfnamefont {B.}~\bibnamefont {Pla\c{c}ais}}, \bibinfo {author}
  {\bibfnamefont {A.}~\bibnamefont {Cavanna}}, \bibinfo {author} {\bibfnamefont
  {Y.}~\bibnamefont {Jin}}, \ and\ \bibinfo {author} {\bibfnamefont
  {G.}~\bibnamefont {F\`eve}},\ }\Doi {10.1126/science.1232572} {\bibfield
  {journal} {\bibinfo  {journal} {Science},\ }\textbf {\bibinfo {volume}
  {339}},\ \bibinfo {pages} {1054} (\bibinfo {year} {2013})}\BibitemShut
  {NoStop}%
\bibitem [{\citenamefont {Dubois}\ \emph {et~al.}(2013)\citenamefont {Dubois},
  \citenamefont {Jullien}, \citenamefont {Roulleau}, \citenamefont {Portier},
  \citenamefont {Roche}, \citenamefont {Cavanna}, \citenamefont {Jin},
  \citenamefont {Wegschneider},\ and\ \citenamefont
  {Glattli}}]{dubois2013nature}%
  \BibitemOpen
  \bibfield  {author} {\bibinfo {author} {\bibfnamefont {J.}~\bibnamefont
  {Dubois}}, \bibinfo {author} {\bibfnamefont {T.}~\bibnamefont {Jullien}},
  \bibinfo {author} {\bibfnamefont {P.}~\bibnamefont {Roulleau}}, \bibinfo
  {author} {\bibfnamefont {F.}~\bibnamefont {Portier}}, \bibinfo {author}
  {\bibfnamefont {P.}~\bibnamefont {Roche}}, \bibinfo {author} {\bibfnamefont
  {A.}~\bibnamefont {Cavanna}}, \bibinfo {author} {\bibfnamefont
  {Y.}~\bibnamefont {Jin}}, \bibinfo {author} {\bibfnamefont {W.}~\bibnamefont
  {Wegschneider}}, \ and\ \bibinfo {author} {\bibfnamefont {D.~C.}\
  \bibnamefont {Glattli}},\ }\Doi {10.1038/nature12713} {\bibfield  {journal}
  {\bibinfo  {journal} {Nature},\ }\textbf {\bibinfo {volume} {502}},\ \bibinfo
  {pages} {659} (\bibinfo {year} {2013})}\BibitemShut {NoStop}%
\bibitem [{\citenamefont {Haake}(2001)}]{Haake2001}%
  \BibitemOpen
  \bibfield  {author} {\bibinfo {author} {\bibfnamefont {F.}~\bibnamefont
  {Haake}},\ }\href@noop {} {\emph {\bibinfo {title} {Quantum Signatures of
  Chaos}}}\ (\bibinfo  {publisher} {Springer-Verlag},\ \bibinfo {year}
  {2001})\BibitemShut {NoStop}%
\bibitem [{\citenamefont {Hassler}\ \emph {et~al.}(2008)\citenamefont
  {Hassler}, \citenamefont {Suslov}, \citenamefont {Graf}, \citenamefont
  {Lebedev}, \citenamefont {Lesovik},\ and\ \citenamefont
  {Blatter}}]{Hassler2008}%
  \BibitemOpen
  \bibfield  {author} {\bibinfo {author} {\bibfnamefont {F.}~\bibnamefont
  {Hassler}}, \bibinfo {author} {\bibfnamefont {M.~V.}\ \bibnamefont {Suslov}},
  \bibinfo {author} {\bibfnamefont {G.~M.}\ \bibnamefont {Graf}}, \bibinfo
  {author} {\bibfnamefont {M.~V.}\ \bibnamefont {Lebedev}}, \bibinfo {author}
  {\bibfnamefont {G.~B.}\ \bibnamefont {Lesovik}}, \ and\ \bibinfo {author}
  {\bibfnamefont {G.}~\bibnamefont {Blatter}},\ }\Doi
  {10.1103/PhysRevB.78.165330} {\bibfield  {journal} {\bibinfo  {journal}
  {Phys. Rev. B},\ }\textbf {\bibinfo {volume} {78}},\ \bibinfo {pages}
  {165330} (\bibinfo {year} {2008})}\BibitemShut {NoStop}%
\bibitem [{\citenamefont {Lesovik}\ and\ \citenamefont
  {Sadovskyy}(2011)}]{Lesovik2011}%
  \BibitemOpen
  \bibfield  {author} {\bibinfo {author} {\bibfnamefont {G.~B.}\ \bibnamefont
  {Lesovik}}\ and\ \bibinfo {author} {\bibfnamefont {I.~A.}\ \bibnamefont
  {Sadovskyy}},\ }\Doi {10.3367/UFNe.0181.201110b.1041} {\bibfield  {journal}
  {\bibinfo  {journal} {Phys. Usp.},\ }\textbf {\bibinfo {volume} {54}},\
  \bibinfo {pages} {1007} (\bibinfo {year} {2011})}\BibitemShut {NoStop}%
\bibitem [{\citenamefont {Choi}\ and\ \citenamefont {Jordan}(2013)}]{Choi2013}%
  \BibitemOpen
  \bibfield  {author} {\bibinfo {author} {\bibfnamefont {Y.}~\bibnamefont
  {Choi}}\ and\ \bibinfo {author} {\bibfnamefont {A.~N.}\ \bibnamefont
  {Jordan}},\ }\Doi {10.1103/PhysRevA.88.052128} {\bibfield  {journal}
  {\bibinfo  {journal} {Phys. Rev. A},\ }\textbf {\bibinfo {volume} {88}},\
  \bibinfo {pages} {052128} (\bibinfo {year} {2013})}\BibitemShut {NoStop}%
\bibitem [{\citenamefont {Haack}\ \emph {et~al.}(2011)\citenamefont {Haack},
  \citenamefont {Moskalets}, \citenamefont {Splettstoesser},\ and\
  \citenamefont {B\"uttiker}}]{Haack2011}%
  \BibitemOpen
  \bibfield  {author} {\bibinfo {author} {\bibfnamefont {G.}~\bibnamefont
  {Haack}}, \bibinfo {author} {\bibfnamefont {M.}~\bibnamefont {Moskalets}},
  \bibinfo {author} {\bibfnamefont {J.}~\bibnamefont {Splettstoesser}}, \ and\
  \bibinfo {author} {\bibfnamefont {M.}~\bibnamefont {B\"uttiker}},\ }\Doi
  {10.1103/PhysRevB.84.081303} {\bibfield  {journal} {\bibinfo  {journal}
  {Phys. Rev. B},\ }\textbf {\bibinfo {volume} {84}},\ \bibinfo {pages}
  {081303(R)} (\bibinfo {year} {2011})}\BibitemShut {NoStop}%
\bibitem [{\citenamefont {Haack}\ \emph {et~al.}(2013)\citenamefont {Haack},
  \citenamefont {Moskalets},\ and\ \citenamefont {B\"uttiker}}]{Haack2013}%
  \BibitemOpen
  \bibfield  {author} {\bibinfo {author} {\bibfnamefont {G.}~\bibnamefont
  {Haack}}, \bibinfo {author} {\bibfnamefont {M.}~\bibnamefont {Moskalets}}, \
  and\ \bibinfo {author} {\bibfnamefont {M.}~\bibnamefont {B\"uttiker}},\ }\Doi
  {10.1103/PhysRevB.87.201302} {\bibfield  {journal} {\bibinfo  {journal}
  {Phys. Rev. B},\ }\textbf {\bibinfo {volume} {87}},\ \bibinfo {pages}
  {201302} (\bibinfo {year} {2013})}\BibitemShut {NoStop}%
\bibitem [{\citenamefont {Buttiker}(1990)\citenamefont {B\"uttiker}}]{Buttiker1990}%
  \BibitemOpen
  \bibfield  {author} {{\bibfnamefont {M.}~\bibnamefont {B\"uttiker}},\ }\Doi
  {10.1103/PhysRevB.41.7906} {\bibfield  {journal} {\bibinfo  {journal}
  {Phys. Rev. B},\ }\textbf {\bibinfo {volume} {41}},\ \bibinfo {pages}
  {7906(R)} (\bibinfo {year} {1990})}\BibitemShut {NoStop}%
\bibitem [{\citenamefont {Landau}\ and\ \citenamefont
  {Lifshitz}(1959)}]{Landau1959}%
  \BibitemOpen
  \bibfield  {author} {\bibinfo {author} {\bibfnamefont {L.~D.}\ \bibnamefont
  {Landau}}\ and\ \bibinfo {author} {\bibfnamefont {E.~M.}\ \bibnamefont
  {Lifshitz}},\ }\href@noop {} {\emph {\bibinfo {title} {Statistical
  Physics}}}\ (\bibinfo  {publisher} {Pergamon, Oxford},\ \bibinfo {year}
  {1959})\BibitemShut {NoStop}%
\bibitem [{\citenamefont {B\"{u}ttiker}(1992)}]{Buttiker1992}%
  \BibitemOpen
  \bibfield  {author} {\bibinfo {author} {\bibfnamefont {M.}~\bibnamefont
  {B\"{u}ttiker}},\ }\Doi {10.1103/PhysRevLett.68.843} {\bibfield  {journal}
  {\bibinfo  {journal} {Phys. Rev. Lett.},\ }\textbf {\bibinfo {volume} {68}},\
  \bibinfo {pages} {843} (\bibinfo {year} {1992})}\BibitemShut {NoStop}%
\bibitem [{\citenamefont {Calogero}(1969)}]{Calogero1969}%
  \BibitemOpen
  \bibfield  {author} {\bibinfo {author} {\bibfnamefont {F.}~\bibnamefont
  {Calogero}},\ }\Doi {http://dx.doi.org/10.1063/1.1664820} {\bibfield
  {journal} {\bibinfo  {journal} {J. Math. Phys.},\ }\textbf {\bibinfo {volume}
  {10}},\ \bibinfo {pages} {2191} (\bibinfo {year} {1969})}\BibitemShut
  {NoStop}%
\bibitem [{\citenamefont {Sutherland}(1971)}]{Sutherland1971}%
  \BibitemOpen
  \bibfield  {author} {\bibinfo {author} {\bibfnamefont {B.}~\bibnamefont
  {Sutherland}},\ }\Doi {http://dx.doi.org/10.1063/1.1665584} {\bibfield
  {journal} {\bibinfo  {journal} {J. Math. Phys.},\ }\textbf {\bibinfo {volume}
  {12}},\ \bibinfo {pages} {246} (\bibinfo {year} {1971})}\BibitemShut
  {NoStop}%
\bibitem [{\citenamefont {Bohigas}\ and\ \citenamefont
  {Pato}(2004)}]{Bohigas2004}%
  \BibitemOpen
  \bibfield  {author} {\bibinfo {author} {\bibfnamefont {O.}~\bibnamefont
  {Bohigas}}\ and\ \bibinfo {author} {\bibfnamefont {M.~P.}\ \bibnamefont
  {Pato}},\ }\Doi {10.1016/j.physletb.2004.05.065} {\bibfield  {journal}
  {\bibinfo  {journal} {Phys. Lett. B},\ }\textbf {\bibinfo {volume} {595}},\
  \bibinfo {pages} {171} (\bibinfo {year} {2004})}\BibitemShut {NoStop}%
\bibitem [{\citenamefont {Bohigas}\ and\ \citenamefont
  {Pato}(2006)}]{Bohigas2006}%
  \BibitemOpen
  \bibfield  {author} {\bibinfo {author} {\bibfnamefont {O.}~\bibnamefont
  {Bohigas}}\ and\ \bibinfo {author} {\bibfnamefont {M.~P.}\ \bibnamefont
  {Pato}},\ }\Doi {10.1103/PhysRevE.74.036212} {\bibfield  {journal} {\bibinfo
  {journal} {Phys. Rev. E},\ }\textbf {\bibinfo {volume} {74}},\ \bibinfo
  {pages} {036212} (\bibinfo {year} {2006})}\BibitemShut {NoStop}%
\bibitem [{\citenamefont {Abanov}(2006)}]{Abanov2006}%
  \BibitemOpen
  \bibfield  {author} {\bibinfo {author} {\bibfnamefont {A.~G.}\ \bibnamefont
  {Abanov}},\ }in\ \href@noop {} {\emph {\bibinfo {booktitle} {Applications of
  Random Matrices in Physics}}},\ \bibinfo {editor} {edited by\ \bibinfo
  {editor} {\bibfnamefont {E.}~\bibnamefont {Brezin}}, \bibinfo {editor}
  {\bibfnamefont {V.}~\bibnamefont {Kazakov}}, \bibinfo {editor} {\bibfnamefont
  {D.}~\bibnamefont {Serban}}, \bibinfo {editor} {\bibfnamefont
  {P.}~\bibnamefont {Wiegmann}}, \ and\ \bibinfo {editor} {\bibfnamefont
  {A.}~\bibnamefont {Zabrodin}}}\ (\bibinfo  {publisher} {Springer},\ \bibinfo
  {year} {2006})\BibitemShut {NoStop}%
\bibitem [{\citenamefont {Wigner}(1955)}]{Wigner1955}%
  \BibitemOpen
  \bibfield  {author} {\bibinfo {author} {\bibfnamefont {E.~P.}\ \bibnamefont
  {Wigner}},\ }\Doi {10.1103/PhysRev.98.145} {\bibfield  {journal} {\bibinfo
  {journal} {Phys. Rev.},\ }\textbf {\bibinfo {volume} {98}},\ \bibinfo {pages}
  {145} (\bibinfo {year} {1955})}\BibitemShut {NoStop}%
\bibitem [{\citenamefont {Smith}(1960)}]{Smith1960}%
  \BibitemOpen
  \bibfield  {author} {\bibinfo {author} {\bibfnamefont {F.~T.}\ \bibnamefont
  {Smith}},\ }\Doi {10.1103/PhysRev.118.349} {\bibfield  {journal} {\bibinfo
  {journal} {Phys. Rev.},\ }\textbf {\bibinfo {volume} {118}},\ \bibinfo
  {pages} {349} (\bibinfo {year} {1960})}\BibitemShut {NoStop}%
\bibitem [{\citenamefont {Ji}\ \emph {et~al.}(2003)\citenamefont {Ji},
  \citenamefont {Chung}, \citenamefont {Sprinzak}, \citenamefont {Heiblum},
  \citenamefont {Mahalu},\ and\ \citenamefont {Shtrikman}}]{Ji2003}%
  \BibitemOpen
  \bibfield  {author} {\bibinfo {author} {\bibfnamefont {Y.}~\bibnamefont
  {Ji}}, \bibinfo {author} {\bibfnamefont {Y.}~\bibnamefont {Chung}}, \bibinfo
  {author} {\bibfnamefont {D.}~\bibnamefont {Sprinzak}}, \bibinfo {author}
  {\bibfnamefont {M.}~\bibnamefont {Heiblum}}, \bibinfo {author} {\bibfnamefont
  {D.}~\bibnamefont {Mahalu}}, \ and\ \bibinfo {author} {\bibfnamefont
  {H.}~\bibnamefont {Shtrikman}},\ }\Doi {10.1038/nature01503} {\bibfield
  {journal} {\bibinfo  {journal} {Nature},\ }\textbf {\bibinfo {volume}
  {422}},\ \bibinfo {pages} {415} (\bibinfo {year} {2003})}\BibitemShut
  {NoStop}%
\bibitem [{\citenamefont {Neder}\ \emph {et~al.}(2006)\citenamefont {Neder},
  \citenamefont {Heiblum}, \citenamefont {Levinson}, \citenamefont {Mahalu},\
  and\ \citenamefont {Umansky}}]{Neder2006}%
  \BibitemOpen
  \bibfield  {author} {\bibinfo {author} {\bibfnamefont {I.}~\bibnamefont
  {Neder}}, \bibinfo {author} {\bibfnamefont {M.}~\bibnamefont {Heiblum}},
  \bibinfo {author} {\bibfnamefont {Y.}~\bibnamefont {Levinson}}, \bibinfo
  {author} {\bibfnamefont {D.}~\bibnamefont {Mahalu}}, \ and\ \bibinfo {author}
  {\bibfnamefont {V.}~\bibnamefont {Umansky}},\ }\Doi
  {10.1103/PhysRevLett.96.016804} {\bibfield  {journal} {\bibinfo  {journal}
  {Phys. Rev. Lett.},\ }\textbf {\bibinfo {volume} {96}},\ \bibinfo {pages}
  {016804} (\bibinfo {year} {2006})}\BibitemShut {NoStop}%
\bibitem [{\citenamefont {Roulleau}\ \emph {et~al.}(2008)\citenamefont
  {Roulleau}, \citenamefont {Portier}, \citenamefont {Roche}, \citenamefont
  {Cavanna}, \citenamefont {Faini}, \citenamefont {Gennser},\ and\
  \citenamefont {Mailly}}]{Roulleau2008}%
  \BibitemOpen
  \bibfield  {author} {\bibinfo {author} {\bibfnamefont {P.}~\bibnamefont
  {Roulleau}}, \bibinfo {author} {\bibfnamefont {F.}~\bibnamefont {Portier}},
  \bibinfo {author} {\bibfnamefont {P.}~\bibnamefont {Roche}}, \bibinfo
  {author} {\bibfnamefont {A.}~\bibnamefont {Cavanna}}, \bibinfo {author}
  {\bibfnamefont {G.}~\bibnamefont {Faini}}, \bibinfo {author} {\bibfnamefont
  {U.}~\bibnamefont {Gennser}}, \ and\ \bibinfo {author} {\bibfnamefont
  {D.}~\bibnamefont {Mailly}},\ }\Doi {10.1103/PhysRevLett.100.126802}
  {\bibfield  {journal} {\bibinfo  {journal} {Phys. Rev. Lett.},\ }\textbf
  {\bibinfo {volume} {100}},\ \bibinfo {pages} {126802} (\bibinfo {year}
  {2008})}\BibitemShut {NoStop}%
\bibitem [{\citenamefont {Litvin}\ \emph {et~al.}(2008)\citenamefont {Litvin},
  \citenamefont {Helzel}, \citenamefont {Tranitz}, \citenamefont
  {Wegscheider},\ and\ \citenamefont {Strunk}}]{Litvin2008}%
  \BibitemOpen
  \bibfield  {author} {\bibinfo {author} {\bibfnamefont {L.~V.}\ \bibnamefont
  {Litvin}}, \bibinfo {author} {\bibfnamefont {A.}~\bibnamefont {Helzel}},
  \bibinfo {author} {\bibfnamefont {H.-P.}\ \bibnamefont {Tranitz}}, \bibinfo
  {author} {\bibfnamefont {W.}~\bibnamefont {Wegscheider}}, \ and\ \bibinfo
  {author} {\bibfnamefont {C.}~\bibnamefont {Strunk}},\ }\Doi
  {10.1103/PhysRevB.78.075303} {\bibfield  {journal} {\bibinfo  {journal}
  {Phys. Rev. B},\ }\textbf {\bibinfo {volume} {78}},\ \bibinfo {pages}
  {075303} (\bibinfo {year} {2008})}\BibitemShut {NoStop}%
\bibitem [{\citenamefont {Bieri}\ \emph {et~al.}(2009)\citenamefont {Bieri},
  \citenamefont {Weiss}, \citenamefont {G\"oktas}, \citenamefont {Hauser},
  \citenamefont {Sch\"onenberger},\ and\ \citenamefont
  {Oberholzer}}]{Bieri2009}%
  \BibitemOpen
  \bibfield  {author} {\bibinfo {author} {\bibfnamefont {E.}~\bibnamefont
  {Bieri}}, \bibinfo {author} {\bibfnamefont {M.}~\bibnamefont {Weiss}},
  \bibinfo {author} {\bibfnamefont {O.}~\bibnamefont {G\"oktas}}, \bibinfo
  {author} {\bibfnamefont {M.}~\bibnamefont {Hauser}}, \bibinfo {author}
  {\bibfnamefont {C.}~\bibnamefont {Sch\"onenberger}}, \ and\ \bibinfo {author}
  {\bibfnamefont {S.}~\bibnamefont {Oberholzer}},\ }\Doi
  {10.1103/PhysRevB.79.245324} {\bibfield  {journal} {\bibinfo  {journal}
  {Phys. Rev. B},\ }\textbf {\bibinfo {volume} {79}},\ \bibinfo {pages}
  {245324} (\bibinfo {year} {2009})}\BibitemShut {NoStop}%
\bibitem [{\citenamefont {Roulleau}\ \emph {et~al.}(2009)\citenamefont
  {Roulleau}, \citenamefont {Portier}, \citenamefont {Roche}, \citenamefont
  {Cavanna}, \citenamefont {Faini}, \citenamefont {Gennser},\ and\
  \citenamefont {Mailly}}]{Roulleau2009}%
  \BibitemOpen
  \bibfield  {author} {\bibinfo {author} {\bibfnamefont {P.}~\bibnamefont
  {Roulleau}}, \bibinfo {author} {\bibfnamefont {F.}~\bibnamefont {Portier}},
  \bibinfo {author} {\bibfnamefont {P.}~\bibnamefont {Roche}}, \bibinfo
  {author} {\bibfnamefont {A.}~\bibnamefont {Cavanna}}, \bibinfo {author}
  {\bibfnamefont {G.}~\bibnamefont {Faini}}, \bibinfo {author} {\bibfnamefont
  {U.}~\bibnamefont {Gennser}}, \ and\ \bibinfo {author} {\bibfnamefont
  {D.}~\bibnamefont {Mailly}},\ }\Doi {10.1103/PhysRevLett.102.236802}
  {\bibfield  {journal} {\bibinfo  {journal} {Phys. Rev. Lett.},\ }\textbf
  {\bibinfo {volume} {102}},\ \bibinfo {pages} {236802} (\bibinfo {year}
  {2009})}\BibitemShut {NoStop}%
\bibitem [{\citenamefont {Thalineau}\ \emph {et~al.}()\citenamefont
  {Thalineau}, \citenamefont {Wieck}, \citenamefont {B\"{a}uerle},\ and\
  \citenamefont {Meunier}}]{Thalineau2014}%
  \BibitemOpen
  \bibfield  {author} {\bibinfo {author} {\bibfnamefont {R.}~\bibnamefont
  {Thalineau}}, \bibinfo {author} {\bibfnamefont {A.~D.}\ \bibnamefont
  {Wieck}}, \bibinfo {author} {\bibfnamefont {C.}~\bibnamefont {B\"{a}uerle}},
  \ and\ \bibinfo {author} {\bibfnamefont {T.}~\bibnamefont {Meunier}},\
  }\href@noop {} {\bibinfo  {journal} {arXiv:1403.7770}}\BibitemShut {NoStop}%
\bibitem [{\citenamefont {Verstraete}\ and\ \citenamefont
  {Cirac}(2010)}]{Verstraete2010}%
  \BibitemOpen
\bibfield  {journal} {  }\bibfield  {author} {\bibinfo {author} {\bibfnamefont
  {F.}~\bibnamefont {Verstraete}}\ and\ \bibinfo {author} {\bibfnamefont
  {J.~I.}\ \bibnamefont {Cirac}},\ }\Doi {10.1103/PhysRevLett.104.190405}
  {\bibfield  {journal} {\bibinfo  {journal} {Phys. Rev. Lett.},\ }\textbf
  {\bibinfo {volume} {104}},\ \bibinfo {pages} {190405} (\bibinfo {year}
  {2010})}\BibitemShut {NoStop}%
\bibitem [{\citenamefont {H\"ubener}\ \emph {et~al.}(2013)\citenamefont
  {H\"ubener}, \citenamefont {Mari},\ and\ \citenamefont
  {Eisert}}]{Hubener2013}%
  \BibitemOpen
  \bibfield  {author} {\bibinfo {author} {\bibfnamefont {R.}~\bibnamefont
  {H\"ubener}}, \bibinfo {author} {\bibfnamefont {A.}~\bibnamefont {Mari}}, \
  and\ \bibinfo {author} {\bibfnamefont {J.}~\bibnamefont {Eisert}},\ }\Doi
  {10.1103/PhysRevLett.110.040401} {\bibfield  {journal} {\bibinfo  {journal}
  {Phys. Rev. Lett.},\ }\textbf {\bibinfo {volume} {110}},\ \bibinfo {pages}
  {040401} (\bibinfo {year} {2013})}\BibitemShut {NoStop}%
\bibitem [{\citenamefont {Mehta}(2004)}]{Mehta2004}%
  \BibitemOpen
  \bibfield  {author} {\bibinfo {author} {\bibfnamefont {M.~L.}\ \bibnamefont
  {Mehta}},\ }\href@noop {} {\emph {\bibinfo {title} {Random Matrices}}}\
  (\bibinfo  {publisher} {Elsevier, Amsterdam},\ \bibinfo {year}
  {2004})\BibitemShut {NoStop}%
\bibitem [{\citenamefont {Emary}\ \emph {et~al.}(2012)\citenamefont {Emary},
  \citenamefont {P\"oltl}, \citenamefont {Carmele}, \citenamefont {Kabuss},
  \citenamefont {Knorr},\ and\ \citenamefont {Brandes}}]{Emary2012}%
  \BibitemOpen
  \bibfield  {author} {\bibinfo {author} {\bibfnamefont {C.}~\bibnamefont
  {Emary}}, \bibinfo {author} {\bibfnamefont {C.}~\bibnamefont {P\"oltl}},
  \bibinfo {author} {\bibfnamefont {A.}~\bibnamefont {Carmele}}, \bibinfo
  {author} {\bibfnamefont {J.}~\bibnamefont {Kabuss}}, \bibinfo {author}
  {\bibfnamefont {A.}~\bibnamefont {Knorr}}, \ and\ \bibinfo {author}
  {\bibfnamefont {T.}~\bibnamefont {Brandes}},\ }\Doi
  {10.1103/PhysRevB.85.165417} {\bibfield  {journal} {\bibinfo  {journal}
  {Phys. Rev. B},\ }\textbf {\bibinfo {volume} {85}},\ \bibinfo {pages}
  {165417} (\bibinfo {year} {2012})}\BibitemShut {NoStop}%
\end{thebibliography}
\end{document}